\pdfoutput=1
\documentclass[%
   final,      
   paper=a4,
   paper=portrait,
   pagesize=auto,
   fontsize=11pt,
   version=last, 
   DIV=calc,
   BCOR=0mm,
   twocolumn
 ]{scrartcl}

\usepackage[T1]{fontenc}
\usepackage[utf8]{inputenc}
\usepackage{textcomp}

\usepackage[english]{babel}
\usepackage[babel, english=british, autostyle=true]{csquotes}

\usepackage[charter,sfscaled=false]{mathdesign}
\usepackage{berasans}
\usepackage{charter}
\usepackage{inconsolata}
\usepackage[scaled]{beramono}

\usepackage[expansion=true,	protrusion=true]{microtype}

\usepackage{relsize}
\usepackage{ragged2e}

\usepackage{graphicx}
\usepackage[table, svgnames]{xcolor}

\definecolor{sectioncolor}{HTML}{53851a}
\definecolor{lightBlue}{RGB}{240, 245, 249}
\definecolor{welfenRed}{RGB}{155, 80, 0}
\definecolor{welfenBlue}{RGB}{0, 80, 155}
\definecolor{welfenGreen}{RGB}{80, 155, 0}
\definecolor{welfenGray}{RGB}{155, 155, 155}

\definecolor{textcolor}{RGB}{0, 0, 0}        
\definecolor{lighttext}{RGB}{128, 128, 128}        

\definecolor{pdfurlcolor}{rgb}{0,0,0.6}
\definecolor{pdffilecolor}{rgb}{0.7,0,0}
\definecolor{pdflinkcolor}{rgb}{0,0,0.6}
\definecolor{pdfcitecolor}{rgb}{0,0,0.6}

\usepackage[centertags, sumlimits, intlimits, namelimits, fleqn]{amsmath} 
\usepackage[fixamsmath,disallowspaces]{mathtools}
\usepackage{fixmath}
\usepackage[all, warning]{onlyamsmath}

\usepackage[Symbolsmallscale]{upgreek}
\usepackage[upmu]{gensymb}  

\usepackage{booktabs}
\usepackage{multirow}
\usepackage{dcolumn}  

\usepackage{threeparttable}
\usepackage{tabularx}

\usepackage{listings}
\usepackage{url} 
\usepackage{enumitem}
\usepackage{fancyhdr}	
\usepackage{lastpage}	

\setkomafont{sectioning}{\sffamily}
\setkomafont{descriptionlabel}{\itshape}
\setkomafont{pageheadfoot}{\normalfont\normalcolor\small\sffamily}
\setkomafont{pagenumber}{\bfseries\usekomafont{sectioning}}
\addtokomafont{sectioning}{\color{sectioncolor}}

\usepackage{titling}
\newcommand{\HorRule}{\color{welfenGray}			
    \rule{\linewidth}{0.5pt}
}
\newcommand{\theabstract}{ABSTRACT GOES HERE}
\newcommand{\setabstract}[1]{\renewcommand{\theabstract}{#1}}

\pretitle{\vspace{0pt} \begin{flushleft} \HorRule \vspace{1mm}\Large \color{sectioncolor} \sffamily \\\vspace{-15pt}}		
\posttitle{\par\end{flushleft}\vskip 0.5em}
\title{Paper title}

\preauthor{\begin{minipage}[t][5.75cm]{0.4\textwidth}\begin{flushleft}\vspace{-8pt}\footnotesize \sffamily \color{sectioncolor}}
\author{Authors}
\postauthor{%
\hspace*{1.3cm}\footnotesize \sffamily
\color{welfenGray}\\[5mm]University of Hannover,\\Welfengarten 1, 30167 Hannover, Germany
\end{flushleft}\vfill\begin{flushleft}
\sffamily\color{sectioncolor}\tiny\textsuperscript{\ding{72}} Corresponding author,\\e-mail: tarnowsky@opentactile.org
\end{flushleft}%
\end{minipage}\hfill\begin{minipage}[t][5.75cm]{0.57\textwidth}
\footnotesize
\textbf{Abstract - }\theabstract
\end{minipage}
\vspace{-5pt}\HorRule\vspace{-30pt}
}

\date{}		

\usepackage[labelfont=bf,textfont=it]{caption}	
\usepackage{fix-cm}
\usepackage[format=plain]{caption}

\usepackage{soul}

\usepackage{siunitx}
\usepackage{multicol}
\usepackage{nicefrac}
\usepackage{glossaries}  
\usepackage{listings}
\usepackage{url}

\usepackage{pifont}
\usepackage{natbib}
\usepackage[switch]{lineno}

\usepackage{float}
\newfloat{lstfloat}{htbp}{lop}
\floatname{lstfloat}{Listing}
\usepackage[absolute]{textpos}

\DeclareMathOperator*{\argmin}{\arg\!\min}

\definecolor{Highlight}{HTML}{69A820}
\definecolor{Gray}{HTML}{AAAAAA}
\definecolor{Keyword}{HTML}{20a3a8}
\definecolor{String}{HTML}{a8a820}

\renewcommand{\c}[1]{\textcolor{Highlight}{\ding{\the\numexpr 201 + #1}}}

\newacronym{i2c}{I\textsuperscript{2}C}{Inter-Integrated Circuit}
\newacronym{spi}{SPI}{Serial Peripheral Interface}
\newacronym{mcu}{MCU}{Main Compute Unit} 
\newacronym{led}{LED}{Light Emitting Diode} 
\newacronym{oled}{OLED}{Organic Light Emitting Diode}
\newacronym{usb}{USB}{Universal Serial Bus}
\newacronym{hid}{HID}{Human Interface Device}
\newacronym{pwm}{PWM}{Pulse Width Modulation}
\newacronym{opamp}{op-amp}{Operational Amplifier}
\newacronym{pcb}{PCB}{Printed Circuit Board}
\newacronym{gpio}{GPIO}{General Purpose Input Output}
\newacronym{ic}{IC}{Integrated Circuit}
\newacronym{pzt}{PZT}{piezoelectric}
\newacronym{hdmi}{HDMI}{High Definition Multimedia Interface}
\newacronym{mos}{MOS}{Metal-Oxide-Semiconductor}
\newacronym{fet}{FET}{Field-Effect Transistor}
\newacronym{dma}{DMA}{Direct Memory Access}
\newacronym{iir}{IIR}{Infinite Impulse Response}
\newacronym{cs}{CS}{Chip Select}
\newacronym{thdn}{THD+N}{Total Harmonic Distortion plus Noise}
\newacronym{ac}{AC}{Alternating Current}
\newacronym{hva}{HVA}{High Voltage Amplification Signal Generator}
\newacronym{asg}{ASG}{Analog Signal Generator}

\title{OpenTactile - An open, modular hardware system\\for controlling tactile displays}
\author{Andreas Tarnowsky\textsuperscript{\ding{72}}, Jan~Jamaszyk, Daniel~Brandes, Franz-Erich~Wolter}

  \setabstract{%
    Tactile displays have a wide potential field of applications, ranging from enhancing Virtual-Reality scenarios up to aiding telesurgery as well as in fundamental psychological and neurophysiological research. In this paper, we describe an open source hardware and software architecture that is designed to drive a variety of different tactile displays.
    For demonstration purposes, a tactile computer mouse featuring a simple tactile display, based on lateral \gls{pzt} actuators, is presented. Even though we will focus on driving mechanical actuators in this paper, the system can be extended to different working principles.
    The suggested architecture is supplied with a custom, easy to use, software stack allowing a simple definition of tactile scenarios as well as user studies while being especially tailored to non-computer scientists.
    By releasing the OpenTactile system under MIT license we hope to ease the burden of controlling tactile displays as well as designing and reproducing the related experiments.}

\begin{document}
\maketitle
\thispagestyle{fancy} 

\begin{textblock*}{2cm}(17.35cm,3.5cm)
\includegraphics[width=2cm]{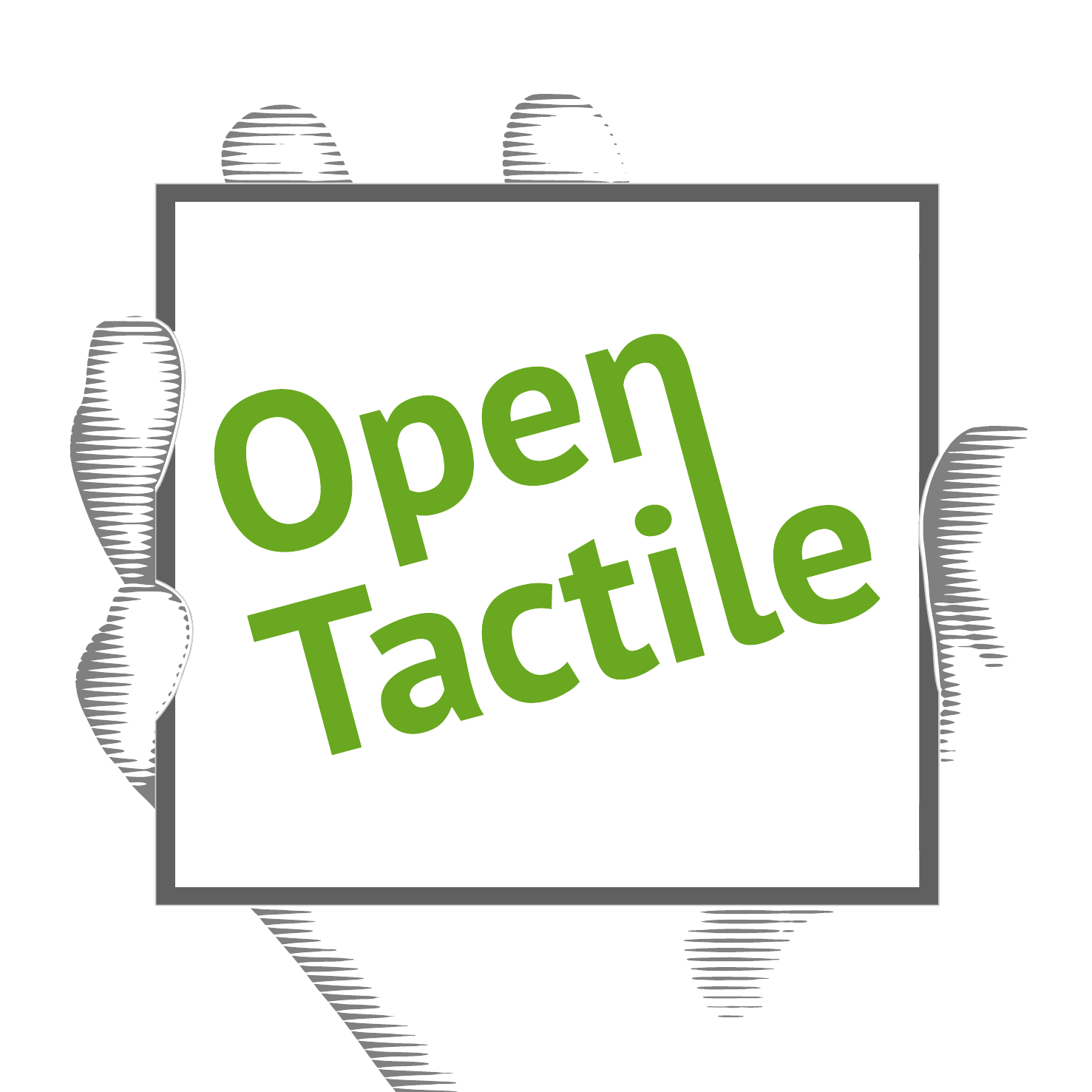}
\end{textblock*}

\section{Introduction}\label{sec:introduction}

While within human computer interaction a realistic generation of arbitrary visual and acoustic impressions has been possible for decades, the question of how to generate realistic tactile impressions still remains open. A key factor for the generation of such impressions are so-called \enquote{tactile displays}, which -- in the mechanical case -- stimulate the user's mechanoreceptive system through the generation of vibratory patterns at the skin surface. To manufacture an actuator, capable of generating realistic impressions researchers examined various types of mechanical principles ranging from microhydraulic actuators [\cite{goethals2008}] to acoustic radiation pressure generated by an array of ultrasound transducers [\cite{Hoshi2010}].

Within the last two decades a plethora of different tactile displays have been developed as has been summarized in [\cite{Benali-Khoudja2004a, ishizuka2015}]. More recent developments broaden the types of mechanical principles even more. For example, a very high resolving -- but not portable -- display has been developed by \emph{Killebrew et al.} [\cite{Killebrew2007}], consisting of a total of $\num{400}$ linear motors. With this high amount of actuators, a spatial resolution of $\SI{0.5}{mm}$ can be achieved on an area of $\SI{1}{cm^2}$.
Other approaches aim for a maximum of portability and flexibility, as for example proposed by \emph{Koo et al.} [\cite{Koo2008}]. Here, an electroactive polymer is driven with high voltages in order to create a displacement perpendicular to the surface of the skin. Even though the spatial resolution of this display is rather low -- just about $\SI{3}{mm}$ --, this display is very thin an can easily be wrapped around e.~g. the finger.
A different way to stimulate the fingertip has been presented by \emph{Wang and Hayward} [\cite{Wang2006}]. While the majority of tactile displays create displacements normal to the skin surface, the so-called \emph{STReSS\textsuperscript{2}} display creates a lateral deformation of the skin by contacting a series of piezoelectric bending actuators directly to the fingertip. With a spatial resolution of $\SI{1.8}{mm}\times\SI{1.2}{mm}$ and a wide temporal bandwidth, this approach allows to create a broad range of tactile impressions.
As \emph{Summers and Chanter} have shown, piezoelectric actuators also can be used to create stimuli acting perpendicular to the skin [\cite{Summers2002}]. Instead of contacting the skin directly, they use L-shaped wire links to redirect the forces to the fingertip. This way, multiple layers of bimorphs can be stacked, resulting in a tactile display with a rather fine resolution of $\SI{1}{mm}\times\SI{1}{mm}$ on an area of $\SI{1}{cm^2}$.
A simplified variant of this principle has been used in the \emph{HAPTEX} project [\cite{Allerkamp2007}]. This display reached a resolution of $\SI{2}{mm}$ by distributing a total of $\num{25}$ actuators over an area of $\SI{1}{cm^2}$. The temporal bandwidth and the maximum amplitudes of this prototype, however, are rather limited.

As different working principles come with the burden of diverse requirements onto the driving hardware, each of the regarded approaches relies on its own hard- as well as software stack, which in most cases are not available to the public. Thus, a reproduction or comparison of the results presented in these publications is hardly possible. Moreover, smaller research teams -- including e.~g. psychologists -- are confronted with a high initial hurdle if they want to start their own experiments using a tactile display.

\subsubsection*{Aim of This Paper}

The goal of this work is to provide an open source hardware and software architecture for driving tactile displays. This hardwaresystem is named \emph{SCRATCHy} -- the SCalable Reference Architecture for a Tactile display Control Hardware. We strive for easing the exchange of hard- and software for tactile displays between different research groups, thus reducing the overhead for conception and development. This way we hope to make tactile displays in general more accessible to other research groups, e.g. psychologists and (neuro-)physiologists, who may not have the expertise in hard- and software architecture. 

By providing a simple yet feature-full programming interface, only basic programming skills are needed to implement own models as well as experimental procedures. The software package also contains a graphical user interface for performing interactive user studies. By incorporating a textual scene description format and a simple programming interface, fairly complex experiments can be realized with a minimum of programming effort.

Alongside the control hardware, we also present a ``tactile mouse'' that fits nicely the proposed hardware-system. It consists of optical sensors to track the position and the orientation of the device as well as a tactile display based on piezoelectric bending actuators. The housing and most of the mechanical parts can easily be obtained via 3D printing.

\subsubsection*{Open Source Policy}
Everything described within this paper will be made publicly available under the MIT-License, allowing everyone to make use of hard- and software, regardless if they support open access or if their work is more restricted. However, we encourage researchers to feed back their improvements to the framework that can be accessed using the following URL:
\\[2mm]
\fbox{\parbox[c][1.0cm]{0.975\columnwidth}{\centering\url{www.opentactile.org}}}\\

We will successively release all data, including:
\begin{itemize}
    \item \gls{pcb} layouts and schematics for all hardware components.    
    \item List of needed electrical components\\i.~e. bill of materials.
    \item Operating software for signal generators as well as libraries to control the host module.
    \item 3D models needed for the creation of a tactile mouse including a tactile display. Firmware for position and orientation sensing.
    \item Additional documentation and tutorials.
\end{itemize}
Up to now the software environment is fully functional, but still under constant development. Therefore we invite everyone to join the project.

\subsection{Structure of this paper}
We subdivided this document into three parts, each one dealing with different aspects of the project.

The first part describes \emph{SCRATCHy}, a hardware architecture that aims for flexibility and scalability in order to drive different kinds of tactile displays. We begin with a short overview of the system architecture, followed by a more detailed description of the hardware- as well as the software-architecture. This section ends with a verification of the system in terms of signal quality and achievable latencies.

In order to offer a more complete tactile system, we introduce in the second part \emph{ITCHy}, a tactile mouse that consists of a tactile display combined with an ergonomic housing and a position sensor. Here we will describe the design of the mechanical components, the electronics needed for tracking the position of the mouse and the software that is needed for USB communication. Again, we will conclude this part with a verification that characterizes the tactile display.

In the last part of this paper, we give closer insights to the software system consisting of various libraries and wrappers, that are intended to make programming the whole system swift and easy. This includes a graphical user interface that is already prepared for carrying out different kinds of user studies. In order to demonstrate how easy the system can be accessed, we will give concrete programming examples.

This paper concludes with a short discussion dealing with possible pitfalls and future improvements to the system.

\section{SCRATCHy -\\ The Control Hardware}
The human skin hosts various types of receptors that respond to mechanical stimuli as well as temperature, itch and pain.
Tactile displays typically try to address three of the mechanoreceptors -- namely the Merkel-, Meissner- and Pacinian-receptors -- that are widely assumed to be responsible for form and texture perception as well as registering movement and distant vibrations [\cite{Johnson2001}]. In short, these three types of receptors and their corresponding afferents cover a large frequency bandwidth, detecting static deflections of the skin as well as vibrations of up to $\SI{1000}{Hz}$ [\cite{Pashler}].
Depending on the specific receptors a tactile display is tuned to, a very high spacial resolution may be needed since Merkel-cells, connected to their corresponding SA1-afferents, are able to resolve spatial details down to a size of $\SI{0.5}{mm}$ [\cite{johnson1981}].

Another aspect to be considered is latency. To put it simple, not only the intensity of a stimulus -- that may be neuronally encoded via the average firing rate of the afferents -- but also the exact timing of neuronal spiking does carry information [\cite{VanRullen2005, Saal2015}]. As pointed out e.~g. in [\cite{Saal2015}], processing of the Pacinian afferents' neuronal signals within the somatosensory cortex may be precise to a millisecond timescale.

To address all these points, we derive the following requirements for the hardware system in order to make it widely applicable to different fields of research:

\begin{itemize}
\item Generation of independent arbitrary stimuli
\item High grade of modularity to allow driving multiple actuators (c.~f. [\cite{Killebrew2007}]: $\num{400}$ actuators on an area of $\SI{1}{cm^2}$).
\item Focus on signal quality in the range between $\SIrange[range-phrase = -]{0}{1000}{Hz}$.
\item Low latencies in order to drive individual actuators within millisecond precision.
\end{itemize}

Especially the requirements of ``modularity'' and ``latency'' are of contrary nature. Therefore we will explicitly analyze to what extend both of them can be fulfilled using our approach. A further important aspect is the flexibility of the system: There are many different principles actuators may be based on. The proposed system should not be limited to a specific concept on generating tactile stimuli.

\subsection{Architecture Overview}
\begin{figure}[b]
\centering
\includegraphics[width=0.9\columnwidth]{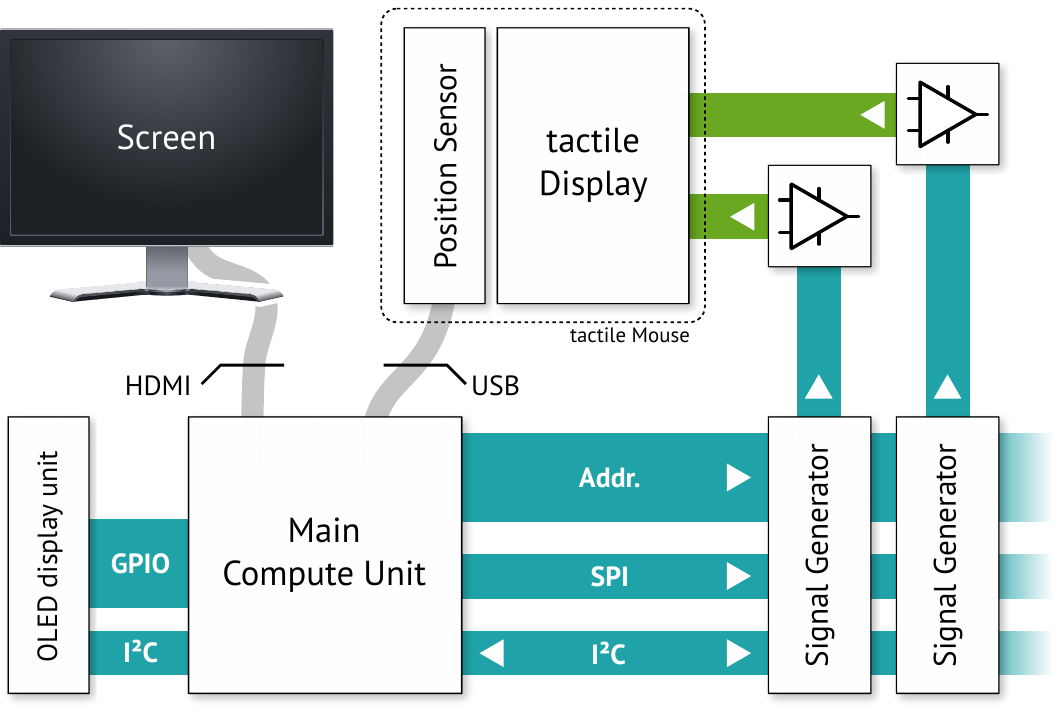}
\caption{Block diagram of the proposed hardware architecture.}
\label{fig:blockdiagram}
\end{figure}
To fulfill the desired low latencies in combination with the flexibility to drive and control various actuator types, we propose a modular multiprocessor architecture. When working with such architectures a key decision needs to be made about the communication protocols as well as its topology. From the need to design an extendable system with respect to the amount of actuators, a natural splitting of the functionality can be derived. A \gls{mcu} is going to be used for the user interaction and model calculation. As an output, an intermediate digital representation of the signal, which should be supplied to an actuator, is generated. This representation is transmitted to a coprocessor using a master/slave architecture. A single coprocessor transforms the intermediate digital representation to up to four signals which, in combination with suiting amplifiers, are used to drive the corresponding actuators. We will denote each coprocessor unit, including any additional electronics for operation, with the term \emph{signal generator} as can also be seen in \emph{figure \ref{fig:blockdiagram}}.

\subsubsection{Data format}
While there is no explicit restriction on the data that can be transferred between the \gls{mcu} and the signal boards, we propose a specific format, that should fit many possible applications and models.
This data format constitutes a \emph{frequency table}, that is a list of tuples each containing frequency and amplitude information. This way the \gls{mcu} is able to work on a more abstract frequency-domain representation of the data, whereas the coprocessors take up to $\num{10}$ such tuples for each actuator to reconstruct a continuous signal within the time-domain.

\subsubsection{Communication protocol}

Regarding the communication between \gls{mcu} and coprocessors several contradictory requirements need to be fulfilled. On the one hand a bidirectional communication is needed for configuration as well as feedback purposes. On the other hand a fast, low latency, unidirectional transmission of the intermediate representation from the \gls{mcu} to the coprocessors is necessary.

Thus, as communication protocols \gls{spi} as well as \gls{i2c} have been chosen. While the \gls{spi} connection -- with speeds over $\SI{10}{Mbit\per s}$ possible -- handles the high speed transfer of the intermediate representation, the \gls{i2c} connection -- typically limited to $\SI{400}{kbit\per s}$ -- is responsible for the slower, but bidirectional communication, as needed e.~g. for configuration and feedback purposes.

Whereas \gls{i2c} comes with a built-in address logic -- which in the classical \SI{7}{bit} mode is limited to \num{112} nodes -- a custom address logic needs to be added for the \gls{spi} communication. In order to prevent any protocol overhead, an additional $\SI{7}{bit}$ \emph{Address-Bus} has been added to communicate with each signal board individually. To assign addresses, a DIP switch in combination with a custom logic takes over the generation of a \gls{cs} signal. Additionally, we added a \emph{Broadcast} signal that can be used to override the address logic and force all signal boards to receive data simultaneously. 

Using a specialized start up procedure, automatically identifying the addresses of all connected coprocessors, the same address logic can be used to assign the \gls{i2c} addresses as well.

\subsubsection{Mechanical considerations}

To easily extend the system a concept of stackable \glspl{pcb} has been developed. All bus systems as well as the power supply are routed through so-called stacking headers. Determining the footprint of the \gls{pcb} it on the one hand needs to be big enough to house the \gls{mcu} and to be flexible with the routing of the traces. On the other hand a small footprint would be beneficial for handling as well as the production costs. Balancing these requirements an area of $\SI{110}{\milli\meter} \times \SI{100}{\milli\meter}$ has been chosen to be fitting.

\subsection{Hardware Architecture}
While the already mentioned \gls{mcu} is the basis for every realization of the proposed hardware-system, the choice of coprocessor boards -- from now on we will denote them as \emph{signal boards} -- is heavily depending on the type of tactile display to be used. In this paper we will present two reference realizations of such a signal board. The first one, the so called \gls{asg} outputs a total of $\num{8}$ analog signals that can be forwarded to an appropriate amplification unit, thus forming a very generic solution. The second board, the \gls{hva}, has been optimized to control piezoelectric bending actuators allowing to drive them directly without the need for external amplification. It can be used e.~g. for driving the ``tactile mouse'' we will describe in \emph{Part II}.

By combining several of these boards, even sophisticated tactile displays can be controlled. In theory, the architecture presented in this paper is able to generate up to $\num{448}$ independent output signals.

\subsubsection{\acrlong{mcu}}
Within the current revision the \gls{mcu} is based on a \emph{Raspberry Pi 3}\footnote{\url{www.raspberrypi.org/products/raspberry-pi-3-model-b/}} embedded system. We have chosen this system, since it is easily available, has a large support community and allows -- thanks to its $\SI{1.2}{GHz}$ quad core CPU -- to implement sophisticated models. 

An optional \emph{display breakout} board enables the usage without any other peripherals (like screen, mouse or keyboard) connected. A small \gls{oled} display, which itself is integrated via \gls{i2c} into the architecture, can show arbitrary information to the user. A configuration of the system -- e.g. a selection of different scenarios -- can be achieved by using the four buttons which are placed on the display breakout as well. For advanced visualization purposes an external screen can be connected to the \gls{hdmi} port of the Raspberry Pi.

\subsubsection{\acrlong{asg}}

The \gls{asg} board can be seen as a reference for the design of a new coprocessor board interfacing the given hardware architecture and thus implements the aforementioned address logic. Since this board does not contain an amplifier stage, the freed-up space has been used to host two separate coprocessor units within one signal board. Therefore, also the address logic has been duplicated -- the address of each coprocessor can be set independently.

In the current revision, we choose a \emph{PJRC Teensy 3.2}\footnote{\url{www.pjrc.com/store/teensy32.html}} USB development board as a coprocessor unit that can be directly mounted to the signal board using a standard multipoint connector. For this kind of application we think it is reasonably priced and offers with its $\SI{32}{bit}$ ARM Cortex-M4 architecture (using a Freescale Semiconductor MK20DX256 processor [\cite{teensy32}]) running at $\SI{96}{MHz}$ enough performance to generate up to four signals simultaneously. Besides performance considerations, it allows -- due to its (partial) Arduino compatibility -- for a very easy programming procedure.

\begin{figure}
\centering
\includegraphics[width=0.9\columnwidth]{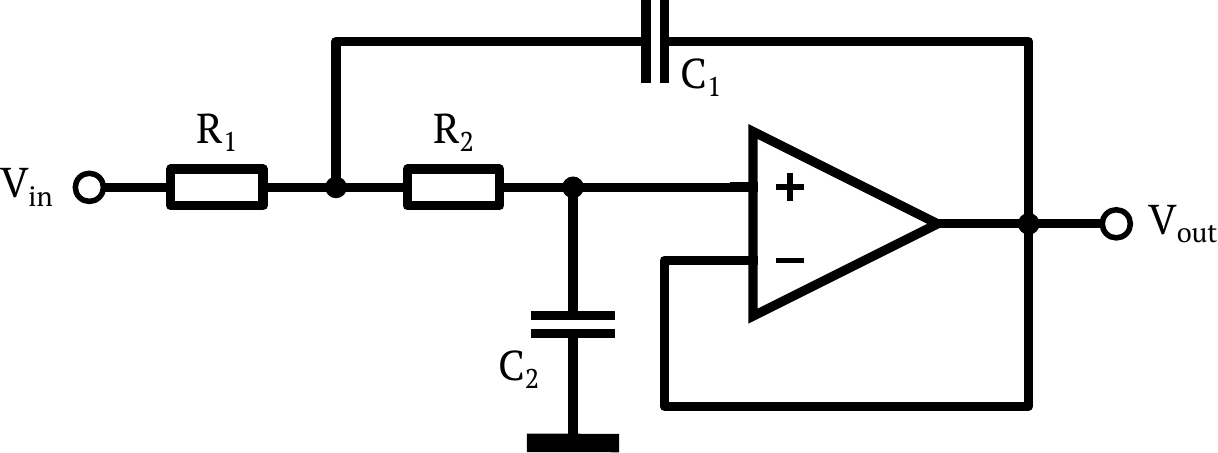}
\caption{Schematic of the used unity gain Sallen-Key low-pass filter topology.}
\label{fig:sallenkey}
\end{figure}
Each of the four generated output signals is digitally encoded using \gls{pwm}. Since an analog output is more appropriate to feed e.~g. external amplifiers, these signals need to be low-pass filtered. This can be achieved for example by using a dedicated filter \gls{ic} as the Maxim Integrated MAX7480. Since production cost have to be taken into account, a less expensive alternative has been realized using \gls{opamp} based low-pass filters.

A standard second order unity gain Sallen-Key topology, as shown in \emph{figure \ref{fig:sallenkey}} consists of four free parameters, namely $C_1$, $C_2$, $R_1$ and $R_2$. To simplify the design process, $C_1$ and $C_2$, which are mainly responsible for the damping described by the Q-factor, are set to $C_1 = \SI{15}{\nano\farad}$ and $C_2 = \SI{10}{\nano\farad}$, resulting in $Q \approx 0.61$. For further simplification $R_1 = R_2 =: R$ can be assumed. Given these simplifications, the undamped natural frequency $\omega_0 = \nicefrac{1}{\sqrt{C_1 C_2} R}$ can be calculated. By specifying the cutoff frequency $f_c$, which as a rule of thumb can be assumed near the natural frequency, the suiting value for $R$ can be estimated:
\begin{linenomath}\begin{align*}
    R &= \frac{1}{2\pi \sqrt{C_1 C_2} f_0} \approx \frac{1}{2\pi \sqrt{C_1 C_2} f_c} \approx \frac{1}{\SI{76.95}{\nano\farad} \cdot f_c}
\end{align*}\end{linenomath}
Regarding the exemplary case of $f_c \approx \SI{1300}{\hertz}$ the resistors can be estimated by $R \approx \SI{10}{\kilo\ohm}$.

A parameter that might be of interest for neurophysiologists, aiming to create specific phase-locked neuronal firing patterns, is the phase shift introduced by the filter. For a calculation, the phase angle of the complex transfer function needs to be calculated. Given the previous Sallen-Key topology the transfer function becomes
\begin{linenomath}\begin{align*}
    H(s) = \frac{1}{1 +2 C_2 R s + C_1 C_2 R^2 s^2} \, ,
\end{align*}\end{linenomath}
thus resulting in a maximum phase shift at $f=\SI{1}{\kilo\hertz}$ of
\begin{linenomath}\begin{align*}
    \varphi(\omega) &= \arg\left(H(j \omega)\right)\\
    \varphi(2\pi\cdot \SI{1}{\kilo\hertz}) &\approx \ang{-72.01}\, .
\end{align*}\end{linenomath}

\subsubsection{\acrlong{hva}}
\begin{figure}
  \centering
  \includegraphics[width=0.9\columnwidth]{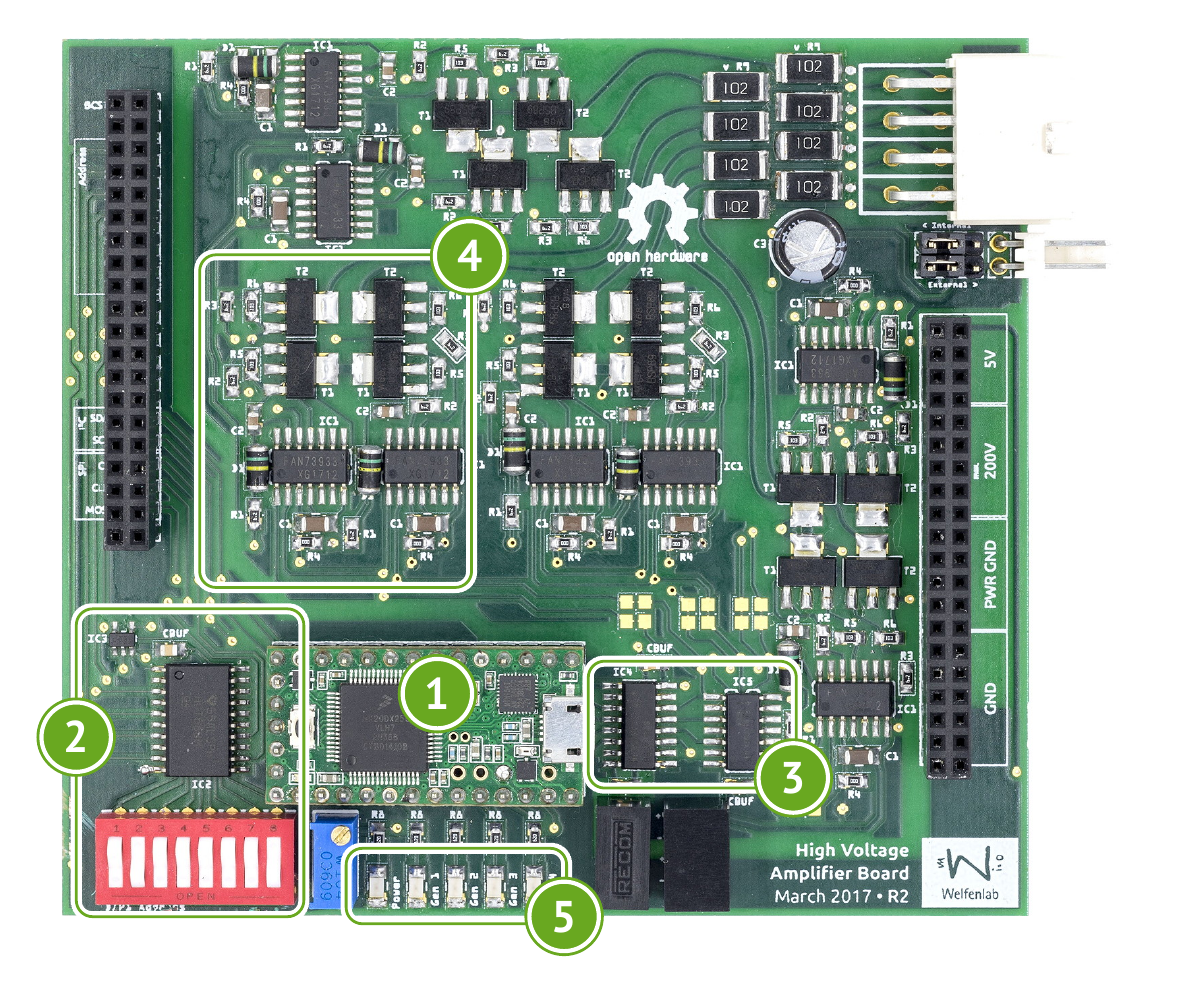}
  \caption{The \gls{hva} board shows the typical structure of a signal board. It contains a processing board \c1 as well as the address logic \c2. The signal is digitally isolated \c3 and then amplified using a total of $\num{4}$ H-bridge drivers \c4. The status of the board is communicated using a series of LEDs \c5.}
  \label{fig:signalboardpcb}
\end{figure}

Especially when actuators based on \gls{pzt} materials are considered for a tactile display, high driving voltages are usually required. As an example, the \gls{pzt} material used in the later on described display is specified for driving voltages of up to $\pm \SI{200}{\volt}$ [\cite{johnsonMattheyBimorph}]. Thus we designed an additional signal board capable of supplying such high voltages by combining a single coprocessor with an isolated $\num{4}$ channel amplification stage. The layout of this board can be seen in \emph{figure \ref{fig:signalboardpcb}}.

For this board we chose a \emph{Class D} amplifier topology that comes with several attributes that are beneficial for this kind of application, including:
\begin{itemize}
    \item \gls{pwm} modulated input,    
    \item Variable supply voltage ($V_{DD}$) only limited\\ by switching transistors and gate driver,
    \item High efficiency,
    \item Symmetrical driving voltage \\by usage of H-bridge topology.
\end{itemize}

\begin{figure}
  \centering
  \includegraphics[width=\columnwidth]{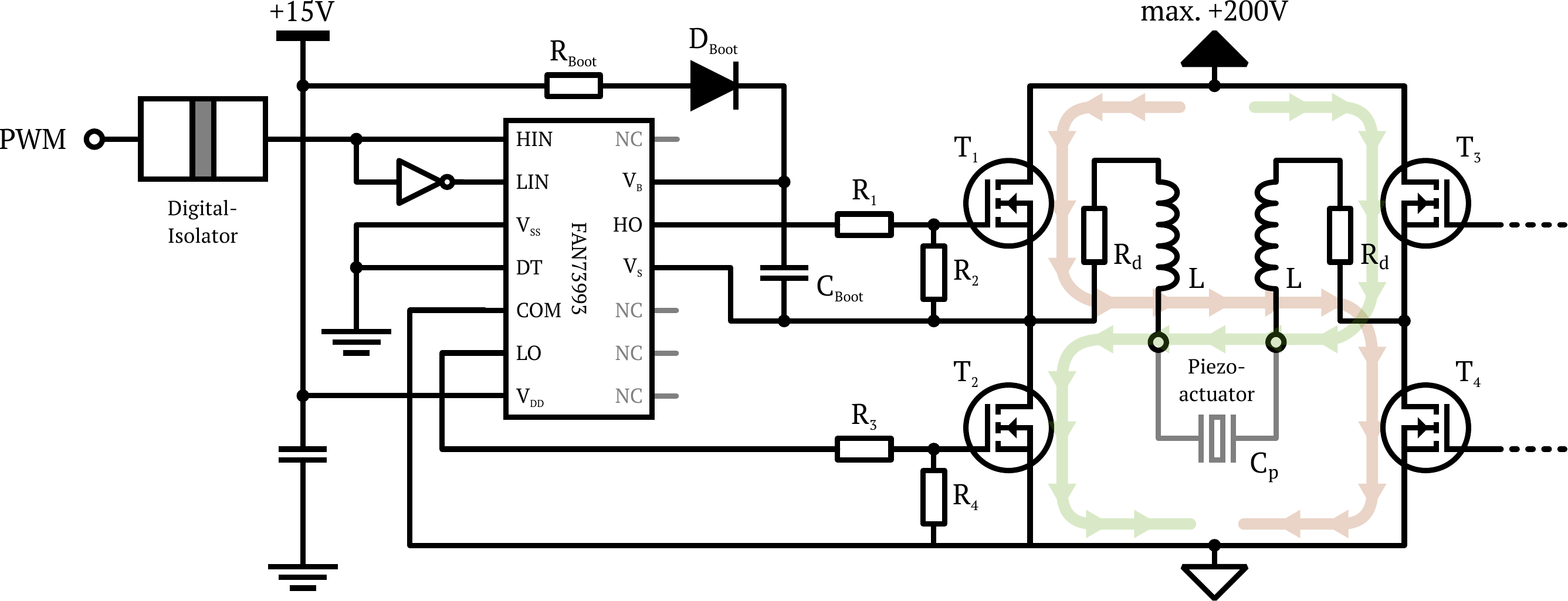}
  \caption{Schematic of a single H-bridge amplifier and the filter topology used to drive a piezoelectric actuator. Each High Voltage Amplifier Board contains four full H-bridge drivers in addition to the processor generating the PWM signals.}
  \label{fig:bridgeschematic}
\end{figure}

The power stage of a \emph{Class-D} amplifier consists of two MOSFET switching transistors. Due to several technical challenges with high power P-MOS transistors a design featuring two N-MOS transistors has been widely accepted as standard. However, this design comes with the burden of generating gate voltages that exceed the supply voltage $V_{DD}$ to open the high side N-MOSFET. To overcome this challenge special \glspl{ic} -- so-called gate drivers -- are available. For the designed \gls{pcb} the FAN73933 gate driver was chosen to be the fitting device. It supports supply voltages up to $V_{DD} = \SI{600}{\volt}$ while still being $\SI{3.3}{\volt}$ and $\SI{5}{\volt}$ compatible on the input side. As power transistors NEXPERIA BSP89, featuring a maximum drain source voltage of \SI{240}{\volt} in combination with a maximum drain current of \SI{340}{\milli\ampere} were chosen. An important component that needs to be optimized to fit the used power transistors is the so-called bootstrap capacitor $C_{BS}$. A thorough optimization, taking into account the charge as well as discharge period of the capacitor led to a selection of $C_{BS} = \SI{10}{\nano\farad}$.

Implementing this topology yields a so called half bridge circuit that comes with a DC offset of $\nicefrac{V_{DD}}{2}$. As for \gls{pzt} actuators a symmetrical driving voltage is beneficial, a so called H-bridge topology has been implemented. The two possible ``current-paths'' that can be generated using such a topology are shown in \emph{figure \ref{fig:bridgeschematic}}. This image also contains one half of the \gls{pcb} schematic, that closely resembles the reference design of the gate driver \gls{ic} [\cite{Fairchild2009}]. For the sake of completeness it should be noted that for different actuator principles a DC offset, as introduced by the half bridge, might even be beneficial. In this case the given amplifier board could be modified to generate eight independent half bridge driven outputs.

The output generated by this topology is a high voltage \gls{pwm} signal and thus still needs to be low-pass filtered. Depending on the requirements regarding the signal quality as well as the used \gls{pwm} modulation scheme various filter topologies, as suggested by [\cite{tiFilter}], may be implemented. For the given amplifier a classical symmetrical RLC filter topology has been chosen. When considering \gls{pzt} actuators their overall capacitive behaviour can actively be exploited in the filter design. As a rule of thumb the cutoff frequency $f_c$ can again be estimated by the natural frequency $f_0$ of the corresponding transfer function. In the given case this can be done by \gls{ac} circuit analysis:
\begin{linenomath}\begin{align*}
    H(s)_{R_d} &= \frac{Z_{C_p}}{Z_g} = \frac{\frac{1}{s C_p}}{\frac{1}{s C_p} +  2 s L + 2 R_L + 2 R_d} \\&= \frac{1}{1 +  2 s^2 L C_p + 2 s (R_L + R_d) C_p}\, ,
\end{align*}\end{linenomath}
which gives $f_0 = \frac{1}{2\pi \sqrt{2LC_p}}$ by comparison. For an exemplary \gls{pzt} actuator with a capacitance of $C_p = \SI{120}{\nano\farad}$ and a desired cutoff frequency of $f_c \approx \SI{1}{\kilo\hertz}$ the filter inductance can be estimated to 
\begin{linenomath}\begin{align*}
L \approx \frac{1}{2 C_p (2\pi f_c)^2} \approx \SI{100}{\milli\henry} \, .
\end{align*}\end{linenomath}
It further needs to be checked if the used inductors are capable of handling the emerging currents without going into saturation. Therefore, a transient simulation of a realistic, high voltage \gls{pwm} signal might be utilized. Coils with such high inductance values tend to come with high serial resistances as well. The used inductor had a serial resistance of $R_L \approx \SI{90}{\ohm}$. Thus, a further damping of the LC circuit might not even be necessary. However, if a higher signal quality and bigger phase shift is preferred over efficiency, an additional damping resistor can be added. An appropriate value for the damping resistor can be obtained by specifying the Q-factor:
\begin{linenomath}\begin{align*}
    Q = \frac{\sqrt{L}}{(R_L + R_d)\sqrt{2 C_p}}
    \Rightarrow R_d = \frac{\sqrt{L}}{Q\sqrt{2 C_p}} - R_L \, .
\end{align*}\end{linenomath}
For an exemplary $Q \approx 0.6$ the damping resistance can be calculated to $R_d \approx \SI{1}{\kilo\ohm}$.

To obtain the maximum phase shift $\varphi$ for $f = \SI{1}{\kilo\hertz}$, as it has been done for the \gls{asg}, the argument transfer function needs to be calculated. Afterwards the maximum phase shift $\varphi_{R_d}(\omega) = \arg\left(H_{R_d}(j\omega)\right)$, with respect to the used damping resistor $R_d$, can be estimated:
\begin{linenomath}\begin{align*}
    \varphi_{\SI{0}{\kilo\ohm}}(2\pi \cdot \SI{1}{\kilo\hertz}) &\approx \ang{-8.89}\\
    \varphi_{\SI{1}{\kilo\ohm}}(2\pi \cdot \SI{1}{\kilo\hertz}) &\approx \ang{-62.16} \, .
\end{align*}\end{linenomath}

The current revision of the board is designed to support supply voltages of up to $\SI{200}{V}$. This limitation arises from the used switching transistors, the bootstrap diode and capacitors as well as the LC-filter components. However, by using appropriate components and a more thorough \gls{pcb} design, taking into account such high driving voltages, the working range could easily be extended to $\SI{600}{\volt}$, which is the maximum specification of the currently used gate driver \gls{ic}. 

\subsection{Software Architecture}
\begin{figure}
\centering
\includegraphics[width=\columnwidth]{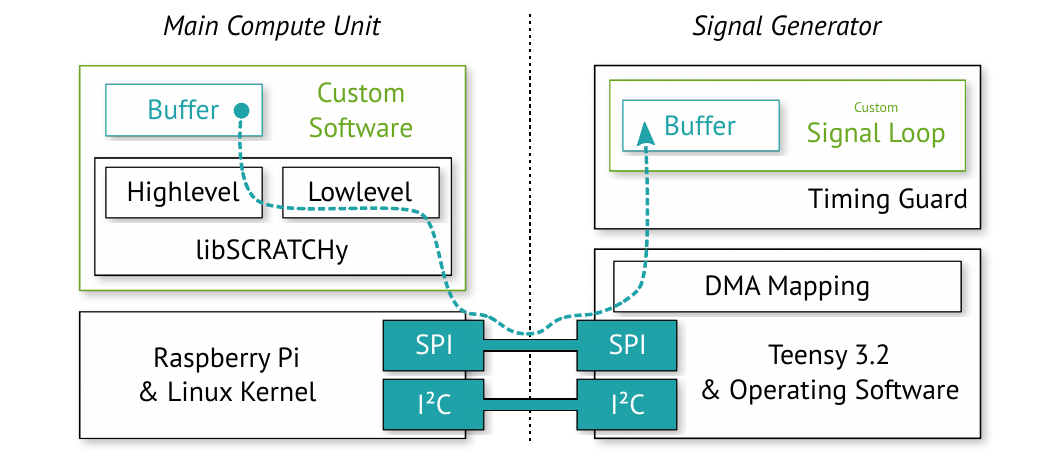}
\caption{Organization of the software architecture. A user defined data structure (buffer) can easily be transmitted via the \gls{spi} bus by utilising the highlevel interface of libSCRATCHy.}
\label{fig:swarchitecture}
\end{figure}

To enable the rapid development and test of new tactile models, the software stack has been designed to provide an easy and intuitive access to the hardware. It consists of several abstraction layers each adding more usability and semantics at the cost of flexibility. A schematic overview over this architecture is depicted in \emph{figure \ref{fig:swarchitecture}}.

Even though the data format used for communication between \gls{mcu} and signal boards can be freely defined, we will focus on the aforementioned \emph{frequency table} approach. We assume that more complex model calculations are carried out within the \gls{mcu} and then expressed as a time-varying frequency table. A stream of such tables is then transferred to the signal boards, that compose the final signal by superimposing all specified frequency components. It should be noted that due to the high update rate via \gls{spi}, parts of the final signal can also be calculated directly within the \gls{mcu}: Since the final signal is reconstructed using cosine functions, the output of the signal generator can be controlled directly by setting the frequency to zero and modulating the amplitude. In this manner low frequency components can be transmitted with unbounded detail whereas higher frequencies still can be mapped to the remaining entries of the frequency table.

\subsubsection{\gls{mcu} Software}

The software stack on the \gls{mcu} consists of a custom designed library (\emph{libSCRATCHy}) written in C++ that handles the coprocessor communication as well as the hardware abstraction. This abstraction allows to easily replace the Raspberry Pi 3 by any other embedded system featuring \gls{spi}, \gls{i2c} and at least twelve \glspl{gpio} without needing to modify higher level components. 

This \emph{Lowlevel-Interface} also enables advanced users to utilize the different components of the system directly. It allows for:

\begin{itemize}
    \item Sending and receiving data via \gls{i2c} and \gls{spi},
    \item Changing the SPI clock frequency, 
    \item Configuring the GPIO ports of the \gls{mcu},
    \item and accessing the OLED display directly.
\end{itemize}

Contrarily, the \emph{Highlevel-Interface} encapsulates these components into a more abstract representation. It consists of four interfaces that are sufficient for most usage scenarios:

\begin{itemize}
\item The \texttt{SignalManager} automatically enumerates all connected signal boards, assigns \gls{i2c} addresses and assures a proper initialization of each board.
\item \texttt{SignalGenerator}s represent individual signal boards and provide an interface to send data via \gls{spi} (e.~g. frequency tables) or to manipulate the internal state-machine directly.
\item The \texttt{GraphicalDisplay} interface allows to output text and small pictograms to the OLED display. It also gives access to the four buttons of the display breakout.
\item Using one of the provided \texttt{PositionQuery} implementations, the position and the current velocity of a connected computer mouse (or the ``tactile mouse'' that additionally provides the actual orientation) can be acquired in terms of absolute coordinates.
\end{itemize}

\subsubsection{Signal Generator Software}
The coprocessors used on all signal boards make use of the so called \emph{Teensyduino} framework and are therefore mostly compatible to the \emph{Arduino} prototyping platform. This allows for a very simple programming procedure taking benefit from the large Arduino community. Additionally, programming can be done via USB, so no additional programming hardware is needed.

In order to simplify the communication with the \gls{mcu}, a basic operating software has been implemented on the coprocessor, taking care -- in conjunction with the \emph{libSCRATCHy} on the \gls{mcu} -- that everything is properly set up during the start up sequence. This software is build upon an internal state-machine that represents different ``runlevels'' of operation. Within these states, different commands can be sent via the \gls{i2c} bus that allow to modify the configuration of each board, including setting the address, changing sampling frequency and the bit-depth of the PWM outputs.

Signal generation is done within a special ``Running'' state, where a single function is repeated indefinitely. This \emph{Signal Loop} can easily be adjusted to fit the needs of the application without demanding to dive into the internals of the system. During operation, the system takes care of proper communication with the \gls{mcu}. For example, data send via the \gls{spi} bus is directly mapped to user accessible memory using \gls{dma} without interrupting the signal loop. This way, the system can guarantee for disruption-free generation of the outgoing signals. To further enforce this guarantee, the system continuously checks if the timing requirements of the configured sampling-rate have been satisfied. In case the timing requirements are not met, the internal state is changed to an ``Error'' state immediately and the user gets noticed.

Exemplarily, the signal generation loop has been implemented to match the aforementioned \emph{frequency table} data format. Within this loop, a total of $\num{40}$ harmonic signals are superimposed to form four PWM outputs, each consisting of $\num{10}$ frequency components. In order to allow for sampling rates as high as $\SI{25}{kHz}$ special care needed to be taken to optimize this loop.

Since the used processor does not include a hardware floating-point unit, the frequency generator has been implemented using the \emph{Q15} fixed point format. Each frequency component is generated using phase accumulation in combination with a $\SI{12}{bit}$ cosine lookup table. The sum of amplitude wise scaled oscillations then forms the final signal. Optionally, all frequency and amplitude values can be smoothed over time to prevent perceivable discontinuities between individual frequency tables. This temporal smoothing is carried out by using two first order \gls{iir} filters for each frequency component, that have been implemented in software.

This example of a possible generator loop forms a ideal starting point for application specific modifications. One could, for example, decrease or increase the number of superimposed wave forms, thus allowing to increase the sampling rate or to create even more complex signals. By adjusting the data format of the frequency table, it would also be easily possible to specify the phase of each signal component. Furthermore, the frequency response of each actuator or possible nonlinearities could be compensated using additional lookup tables.

\subsection{Verification and Performance Evaluation}
Albeit we already use this system for operating different kinds of tactile displays successfully, we want to give additional characterizations of the performance this architecture can provide. Here we will especially focus on the signal quality permitted by the signal boards as well as the average latencies that can be expected in a productive scenario.

To address the scalability of the architecture, we tested the system on a novel bimodal tactile display consisting of $\num{16}$ piezoelectric bending actuators that were combined with another $\num{16}$ electromagnetic actuators, thus bearing a total of $\num{32}$ degrees of freedom [\cite{dfg}]. While the \gls{pzt} actuators were driven by four \gls{hva} boards, the electromagnetical part of the display used the amplified signal of two \gls{asg} boards. In this configuration, the system allowed for stable operation at any time.

\subsubsection{Signal Quality}

\begin{table}[p]
  \centering
  \begin{tabularx}{\columnwidth}{p{1.4cm}p{1.15cm}p{0.1cm}p{1cm}p{1cm}p{1cm}p{1cm}}  
    \toprule
    Frequency {\scriptsize(Target)} & Frequency {\scriptsize(Measured)} & &THD+N \SI{1}{kHz} & THD+N \SI{20}{kHz} & THD+N \SI{1}{MHz} \\
    \midrule
	\SI{10}{Hz} & \SI{9.92}{Hz} & &\cellcolor{Highlight!25}\SI{0.14}{\%} & \SI{0.16}{\%} & \SI{0.68}{\%} \\
	\SI{50}{Hz} & \SI{49.59}{Hz} & &\cellcolor{Highlight!25}\SI{0.23}{\%} & \SI{0.24}{\%} & \SI{0.27}{\%} \\
	\SI{125}{Hz} & \SI{124.36}{Hz} & &\cellcolor{Highlight!25}\SI{0.27}{\%} & \SI{0.28}{\%} & \SI{0.31}{\%} \\
	\SI{250}{Hz} & \SI{249.47}{Hz} & &\cellcolor{Highlight!25}\SI{0.37}{\%} & \SI{0.39}{\%} & \SI{0.40}{\%} \\
	\SI{500}{Hz} & \SI{499.70}{Hz} & &\cellcolor{Highlight!25}\SI{0.22}{\%} & \SI{0.45}{\%} & \SI{0.46}{\%} \\
	\SI{750}{Hz} & \SI{749.90}{Hz} & &\cellcolor{Highlight!25}\SI{0.05}{\%} & \SI{0.50}{\%} & \SI{0.51}{\%} \\
	\SI{1000}{Hz} & \SI{999.34}{Hz} & &\cellcolor{Highlight!25}\SI{0.02}{\%} & \SI{0.57}{\%} & \SI{0.59}{\%} \\
    \bottomrule
\end{tabularx}
\caption{Measurements of THD+N for the \gls{asg} board. Values relevant to tactile perception are highlighted.}
\label{tab:THDN_ASG}
\end{table}
\begin{table}[p]
\centering
    \begin{tabularx}{\columnwidth}{p{1.4cm}p{1.15cm}p{0.1cm}p{1cm}p{1cm}p{1cm}p{1cm}}  
    \toprule
    Frequency {\scriptsize(Target)} & Frequency {\scriptsize(Measured)} & &THD+N \SI{1}{kHz} & THD+N \SI{20}{kHz} & THD+N \SI{1}{MHz} \\
    \midrule
	\SI{10}{Hz} & \SI{10.68}{Hz} & &\cellcolor{Highlight!25}\SI{0.27}{\%} & \SI{0.31}{\%} & \SI{0.63}{\%} \\
	\SI{50}{Hz} & \SI{50.35}{Hz} & &\cellcolor{Highlight!25}\SI{0.62}{\%} & \SI{0.65}{\%} & \SI{0.71}{\%} \\
	\SI{125}{Hz} & \SI{125.88}{Hz} & &\cellcolor{Highlight!25}\SI{1.01}{\%} & \SI{1.03}{\%} & \SI{1.06}{\%} \\
	\SI{250}{Hz} & \SI{249.47}{Hz} & &\cellcolor{Highlight!25}\SI{1.27}{\%} & \SI{1.38}{\%} & \SI{1.42}{\%} \\
	\SI{500}{Hz} & \SI{499.68}{Hz} & &\cellcolor{Highlight!25}\SI{0.57}{\%} & \SI{1.42}{\%} & \SI{1.46}{\%} \\
	\SI{750}{Hz} & \SI{749.89}{Hz} & &\cellcolor{Highlight!25}\SI{0.56}{\%} & \SI{1.62}{\%} & \SI{1.69}{\%} \\
	\SI{1000}{Hz} & \SI{998.86}{Hz} & &\cellcolor{Highlight!25}\SI{0.79}{\%} & \SI{1.80}{\%} & \SI{1.89}{\%} \\
    \bottomrule
    \end{tabularx}
    \caption{Measurements of THD+N for the \gls{hva} board. Values relevant to tactile perception are highlighted.}
    \label{tab:THDN_HVA}
\end{table}
\begin{table}[p]
\centering
\begin{tabularx}{\columnwidth}{p{1.5cm}p{1.25cm}p{1.25cm}p{1.25cm}p{1.75cm}}  
    \toprule
    SPI Freq. & $t_\text{Data}^1$  & $t_\text{Data}^2$ & $t_\text{Data}^8$ & Data rate \\
    \midrule
	\SI{967}{kHz} & \SI{1480}{\mu s} & \SI{2960}{\mu s} & \SI{11840}{\mu s} & \SI{108}{kB\per\second} \\
	\SI{1953}{kHz} & \cellcolor{Highlight!25}\SI{742}{\mu s} & \SI{1484}{\mu s} & \SI{5931}{\mu s} & \SI{216}{kB\per\second} \\
	\SI{3.9}{MHz} & \cellcolor{Highlight!25}\SI{374}{\mu s} & \cellcolor{Highlight!25}\SI{746}{\mu s} & \SI{2980}{\mu s} & \SI{430}{kB\per\second} \\
	\SI{7.8}{MHz} & \cellcolor{Highlight!25}\SI{190}{\mu s} & \cellcolor{Highlight!25}\SI{377}{\mu s} & \SI{1503}{\mu s} & \SI{852}{kB\per\second} \\
	\SI{15.6}{MHz} & \cellcolor{Highlight!25}\SI{98}{\mu s} & \cellcolor{Highlight!25}\SI{196}{\mu s} & \cellcolor{Highlight!25}\SI{779}{\mu s} & \SI{1643}{kB\per\second} \\
    \bottomrule
\end{tabularx}
\caption{Time needed for transmitting a data package of $\SI{40}{bytes}$ to each of $1$, $2$ and $8$ signal boards for various \gls{spi} clock frequencies. Values satisfying the $\SI{1000}{Hz}$ latency goal are highlighted.}
\label{tab:Latency}
\end{table}

To quantify the signal quality of both kinds of signal boards objectively, we measured the so-called \emph{\gls{thdn}} for various frequencies ranging from \SI{10}{Hz} to \SI{1000}{Hz}. The \gls{thdn} is defined as (c.~f. [\cite{Lai2009}])
\begin{linenomath}\begin{align*}
    \text{THD+N}_\% = \frac{100}{U_1} \sqrt{\sum_n U_n^2 + U_\text{noise}^2}\,, 
\end{align*}\end{linenomath}
with $U_1$ being the root mean square voltage of the signal, $U_n$ its harmonics and $U_\text{noise}$ the non harmonic distortion of the signal respectively. It can be easily interpreted as the amount of nonlinearities distorting the original signal. 

In practice, we firstly measured the harmonic signals with its respective frequencies outputted by each board using a \emph{Rigol DS2072A} oscilloscope, resulting in a set of discretely sampled values $\hat{u}_i$ with a fixed samplingwidth of $\Delta t = \SI{5e-6}{s}$. We then fit an analytical sine function to this data by numerical minimization as follows:
\begin{linenomath}\begin{align*}
    \argmin\limits_{A, f, \varphi \in \mathbb{R}} \underbrace{\sum\limits_{i = 0}^N \left\lVert \hat{u}_{i} - A\cdot\sin(2\pi f \cdot i \Delta t + \varphi) \right\rVert^2}_{u_{\text{Noise}}^2}\,.
\end{align*}\end{linenomath}
The residuum $u_{\text{Noise}}^2$ of this minimization process then can be directly used to calculate the \gls{thdn} of the signal:
\begin{linenomath}\begin{align*}
    \text{THD+N}_\% = \frac{100}{A} \sqrt{\frac{u_{\text{Noise}}^2}{N}}\,.
\end{align*}\end{linenomath}
By measuring the \gls{thdn} this way, we are able to address all kinds of signal degradations that may emerge from nonlinearities within the hardware system as well as e.~g. quantization errors due to the fixed point format used in the signal generation process. Additional, we can compare the targeted frequency of the signal versus the actually generated waveform.

The actual measured \gls{thdn} values for both signal boards presented here are given in \emph{table \ref{tab:THDN_ASG}} and \emph{table \ref{tab:THDN_HVA}} respectively. By low-pass filtering the noise term, the influence of signal disturbances within the frequency range of tactile perception (THD+N \SI{1}{kHz}) as well as acoustic perception (THD+N \SI{20}{kHz}) can be derived. Technically, the signal of the \gls{asg} board has been measured in the unloaded case, whereas the \gls{hva} board was connected to the tactile display described in Part II. In a worst case scenario, the amplified signal of the \gls{hva} may contain deviations of about \SI{1.27}{\%} of the driving voltage. Considering the characteristics of the tactile display presented in the next part of this paper, this error roughly translates to a maximum amplitude of $< \SI{0.54}{\mu m}$ in the loaded case ($\pm \SI{60}{V}$) that we assume to be uncritical.

\subsubsection{System Latency}
The second series of measurements addresses the data throughput that can be achieved using the unidirectional \gls{spi} protocol used for distributing e.~g. the frequency tables. These values are directly linked to the latencies of the system one could expect.

In order to measure the time span needed to send data packages with an exemplary size of $\SI{40}{bytes}$, we utilize an additional \gls{gpio} output of the Raspberry Pi that is used to accurately track the time needed for program execution and sending data. Within the main loop running on the \gls{mcu}, this output is set to a high state before \gls{spi} communication is initiated and set back to a low state after execution of the \gls{spi} commands finished. Since the data is transferred to the signal boards using \gls{dma} transfers, the changes are processed immediately and additional delays depend solely on the software implementation of the coprocessors. In our concrete case, this delay depends on the chosen sampling rate of the signal generation procedure as well as the digital \gls{iir} filtering stage. After measuring this delay to be in the range of few microseconds, we found it to be negligible. The total time needed for such a loop iteration then is tracked by the aforementioned oscilloscope. For each measurement the \gls{spi} clock frequency was varied and the time for sending the table was averaged across at least $\num{500}$ loop iterations. This procedure was repeated for different configurations of the system consisting of $1$, $2$ and $8$ signal boards.

The resulting timing information is presented in \emph{table \ref{tab:Latency}}. As expected, the time needed for sending the frequency tables depends approximately linear on the number of signal boards and the \gls{spi} clock speed. In this scenario, a total of $32$ actuators can be accessed within $\SI{779}{\mu s}$ leaving about $\SI{220}{\mu s}$ for model calculations and user interaction. In case such high update rates are needed for higher resolution displays, one can simply reduce the amount of frequency components per table, thus reducing the required bandwidth to transmit the data packages.

\section{ITCHy - The Tactile Mouse}
Besides the SCRATCHy hardwaresystem, we also provide the schematics, electronics and 3D models for manufacturing a ``tactile mouse'' that
enables the user to freely explore tactile scenarios. This device -- we call it the Interactive Tactile display Control Handle (ITCHy) -- contains two laser sensors to track the position, orientation and current movement speed of the mouse, as well as a tactile display that is based on \gls{pzt} actuators. 

Since different displays based on the \gls{pzt} principle already exist, our goal was to create a display that achieves competitive performance while being easy to manufacture and assemble. The resulting system mounts piezoelectric actuators using a simple clamping mechanism and comes in an ergonomic housing that closely resembles a conventional computer mouse.

\subsection{The Tactile Mouse}
\begin{figure*}
\centering
\includegraphics[height=4.5cm]{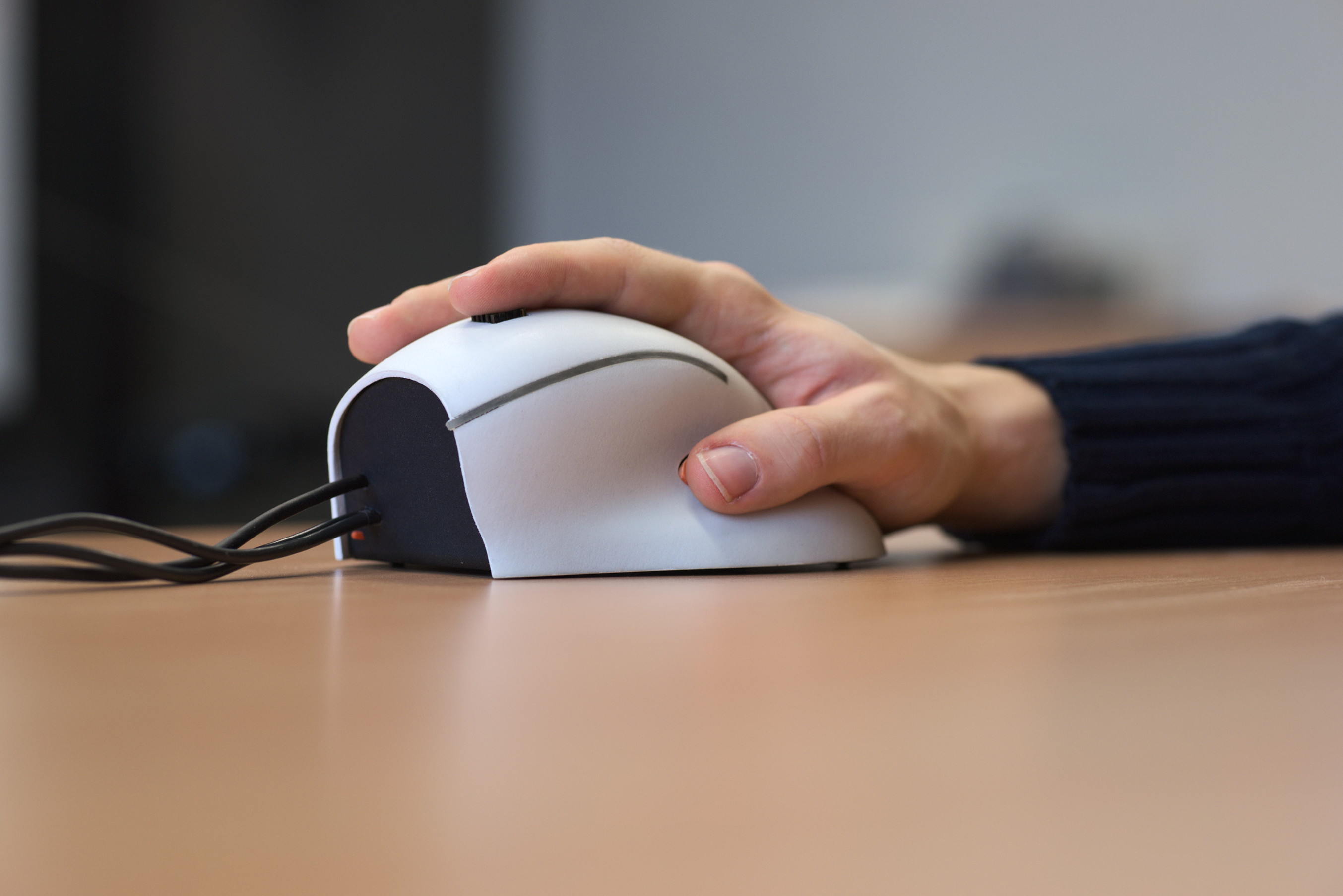}\hspace{1mm}\includegraphics[height=4.5cm]{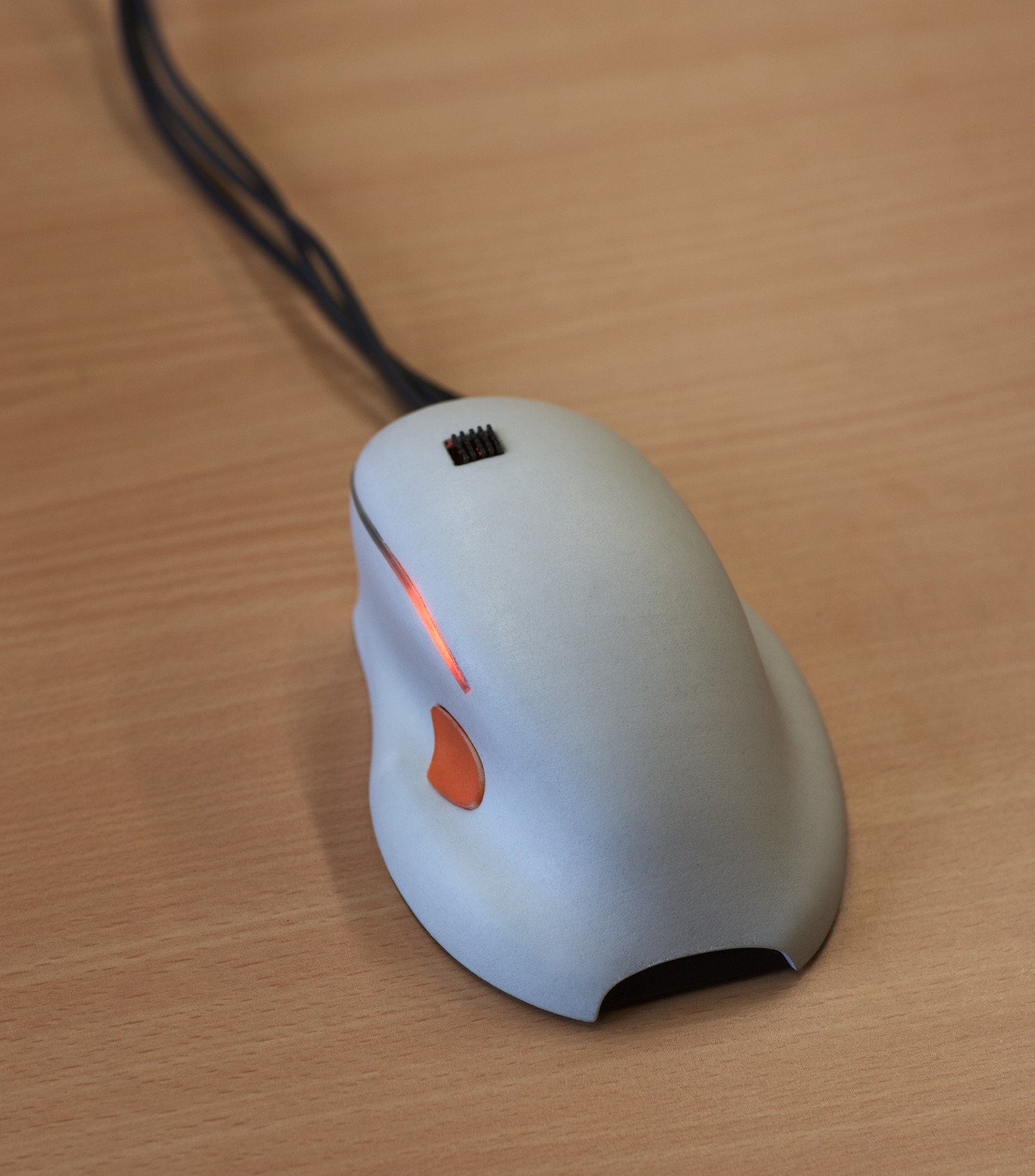}\hspace{1mm}\includegraphics[height=4.5cm]{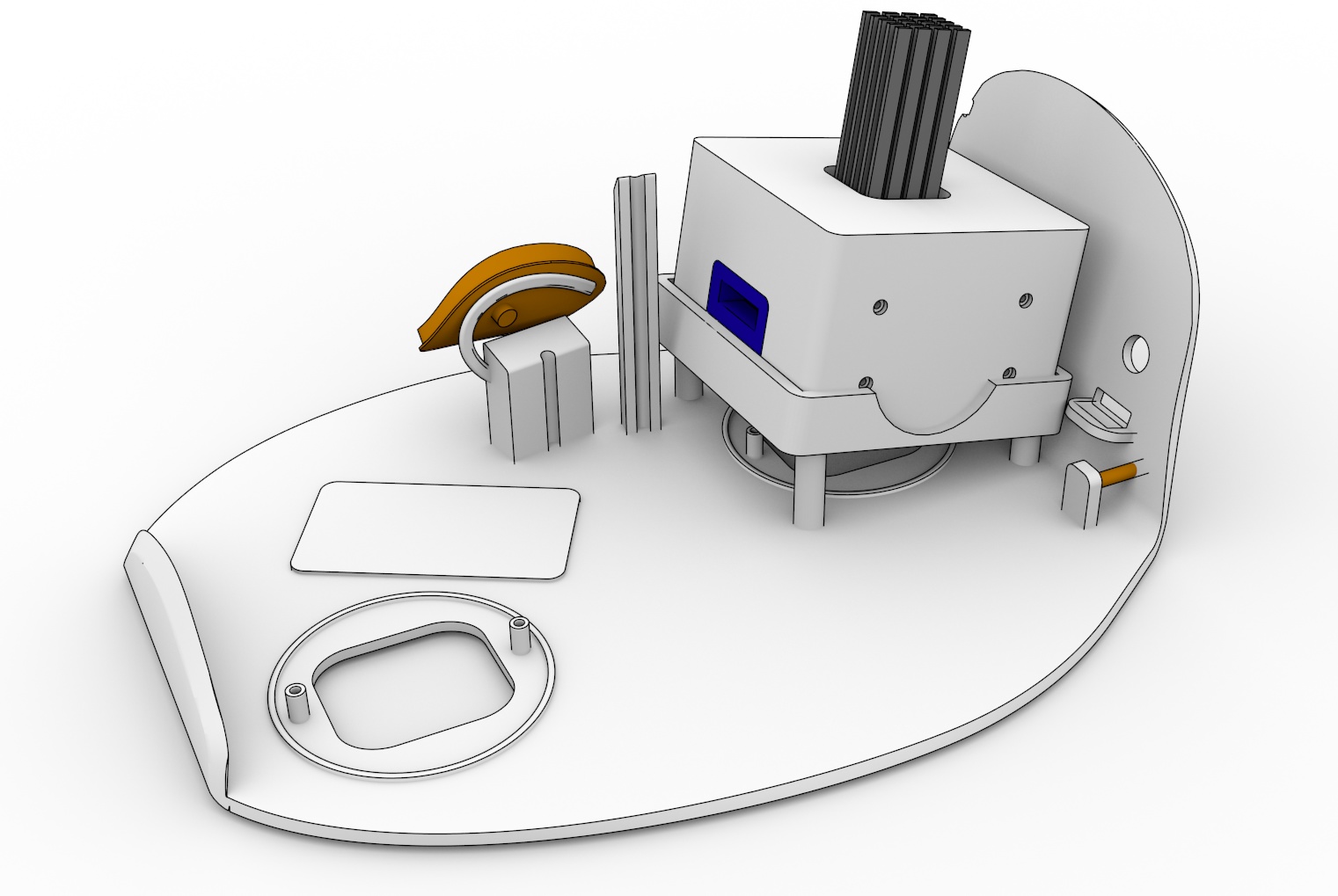}
\caption{Photographs of the assembled tactile mouse. In the middle image, the integrated LED as well as a thumb button are visible. As can be barely seen on the left image, an additional small button on the front can be used to reset the orientation and position of the device. The right image shows a rendering of the internal structure.}
\label{fig:tactilemouse}
\end{figure*}
The housing of the tactile mouse consist of five separate parts, that can be 3D printed using standard processes that allow for a precision of approximately $\pm\SI{0.2}{mm}$.\footnote{Our concrete prototype was manufactured by \emph{Shapeways} using the \emph{Strong \& Flexible Plastic} as well as transparent \emph{Acrylic Plastic}.} Most of these parts can be identified within \emph{figure \ref{fig:tactilemouse}}: Two smaller elements form buttons (orange) that are used for user interaction. The (white) hand rest is ergonomically formed and contains a breach that fits the actuators of the tactile display. It also mounts a small diffuser made from transparent plastic that scatters the light of a built-in RGB \gls{led}. The base element (black) of the mouse comprises a mount for the tactile display, that is tilted in a way such that the actuators follow the contour of the hand rest. Additionally, the bottom part contains a stencil for aligning the optical sensors.

\subsection{The Tactile Display}

In order to test the whole system, we developed a relatively simple tactile display, that induces lateral motion to the fingertip, using a similar principle as has already been seen in [\cite{Wang2006}]. It consists of a total of $\num{20}$ ``Type 2'' piezoelectric bending actuators manufactured by \emph{Johnson Matthey Piezo Products} [\cite{johnsonMattheyBimorph}]. Due to its high blocking force and large maximum deflection, the piezoceramic material ``VIBRIT M1876'' has been chosen [\cite{Matthey2014}]. 

\begin{figure*}
\centering
\includegraphics[height=3.5cm]{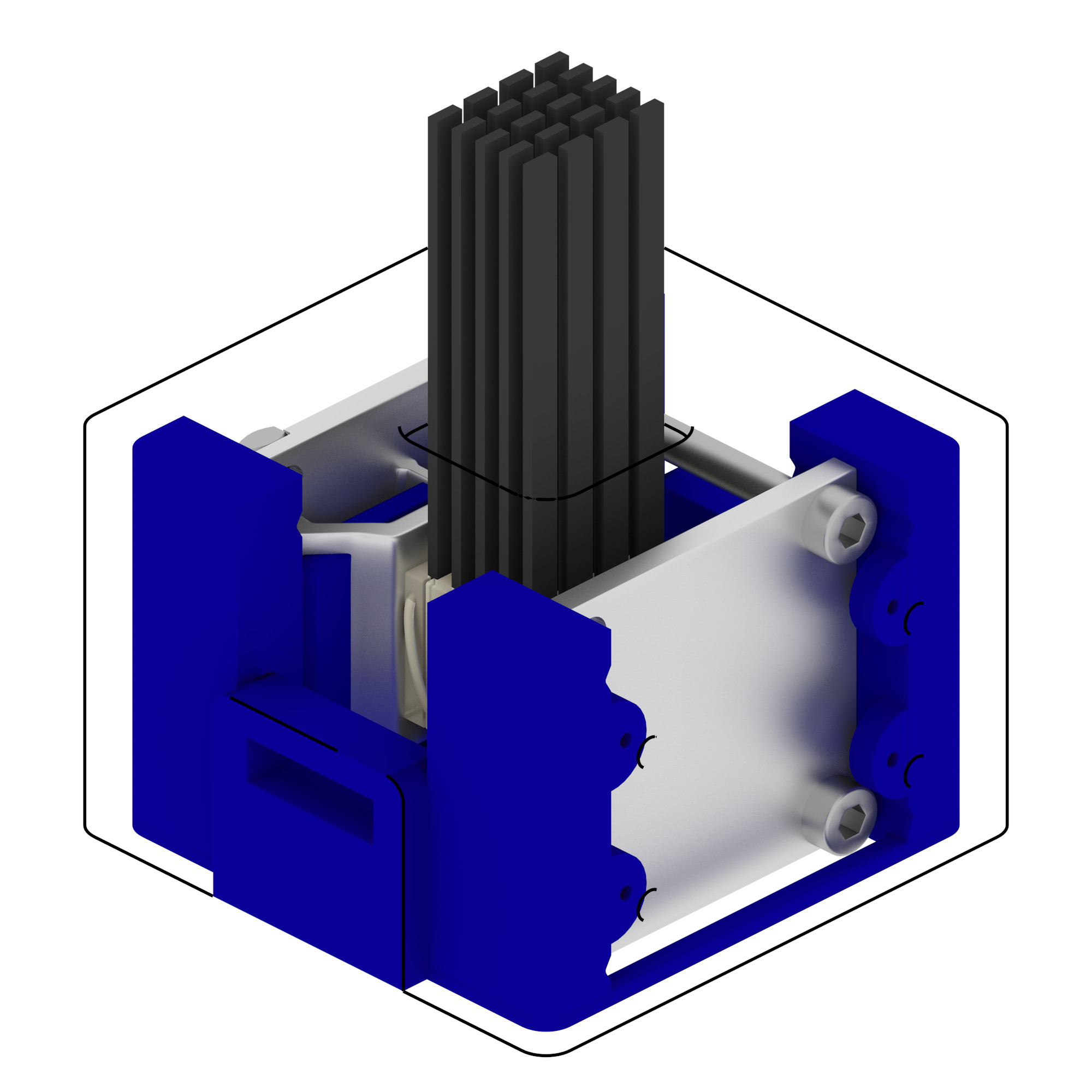}\hfill\includegraphics[height=3.5cm]{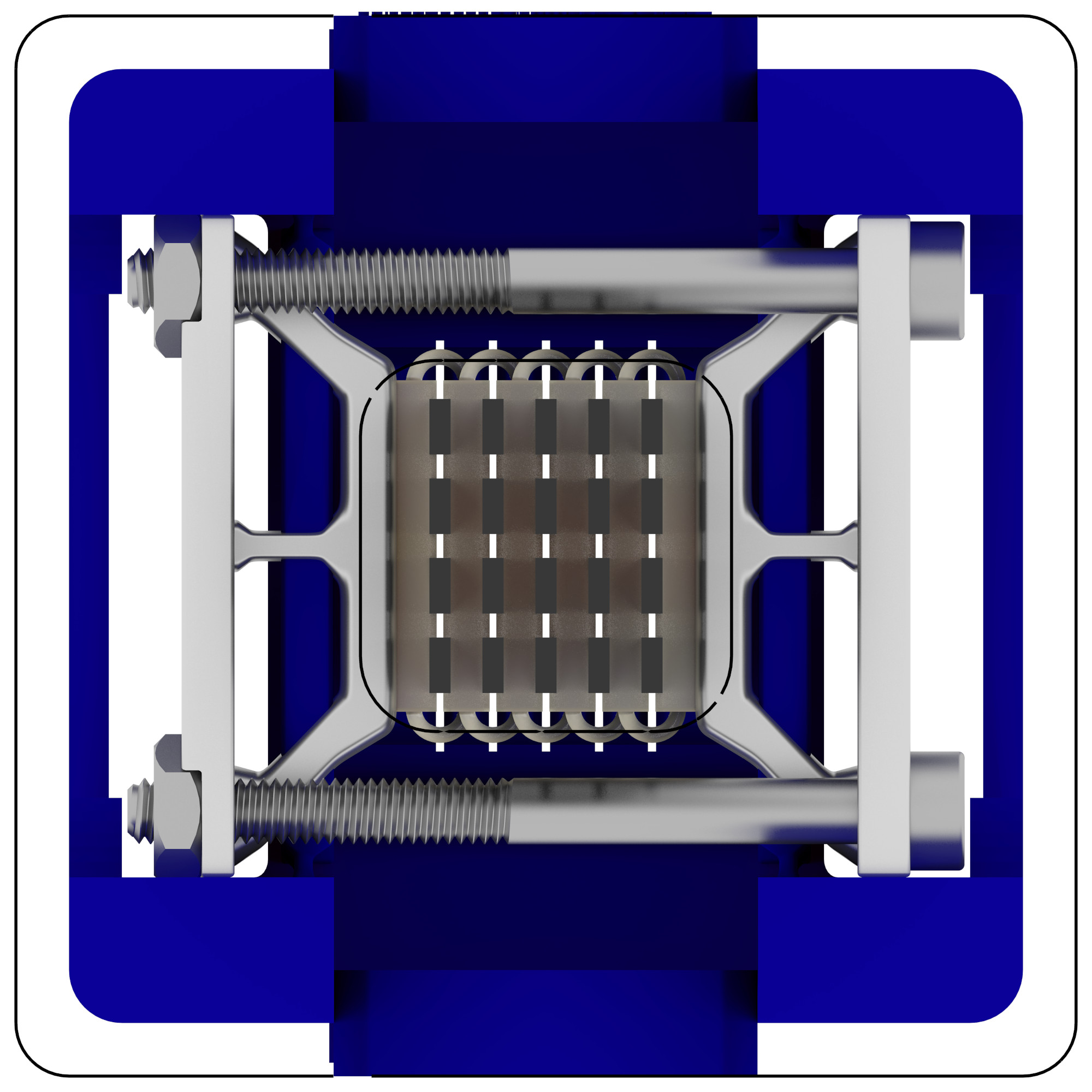}\hfill\includegraphics[height=3.5cm]{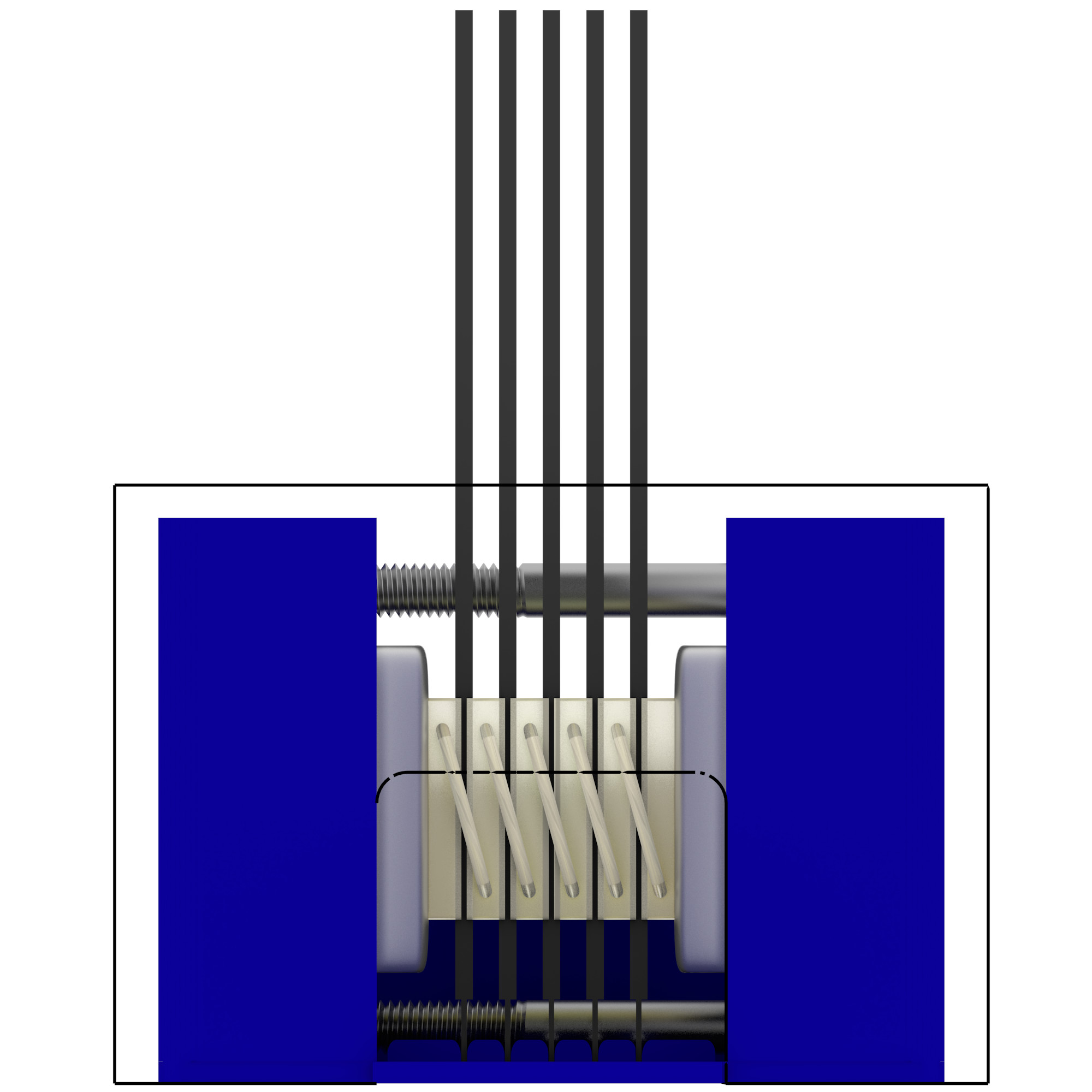}\hfill\includegraphics[height=3.5cm]{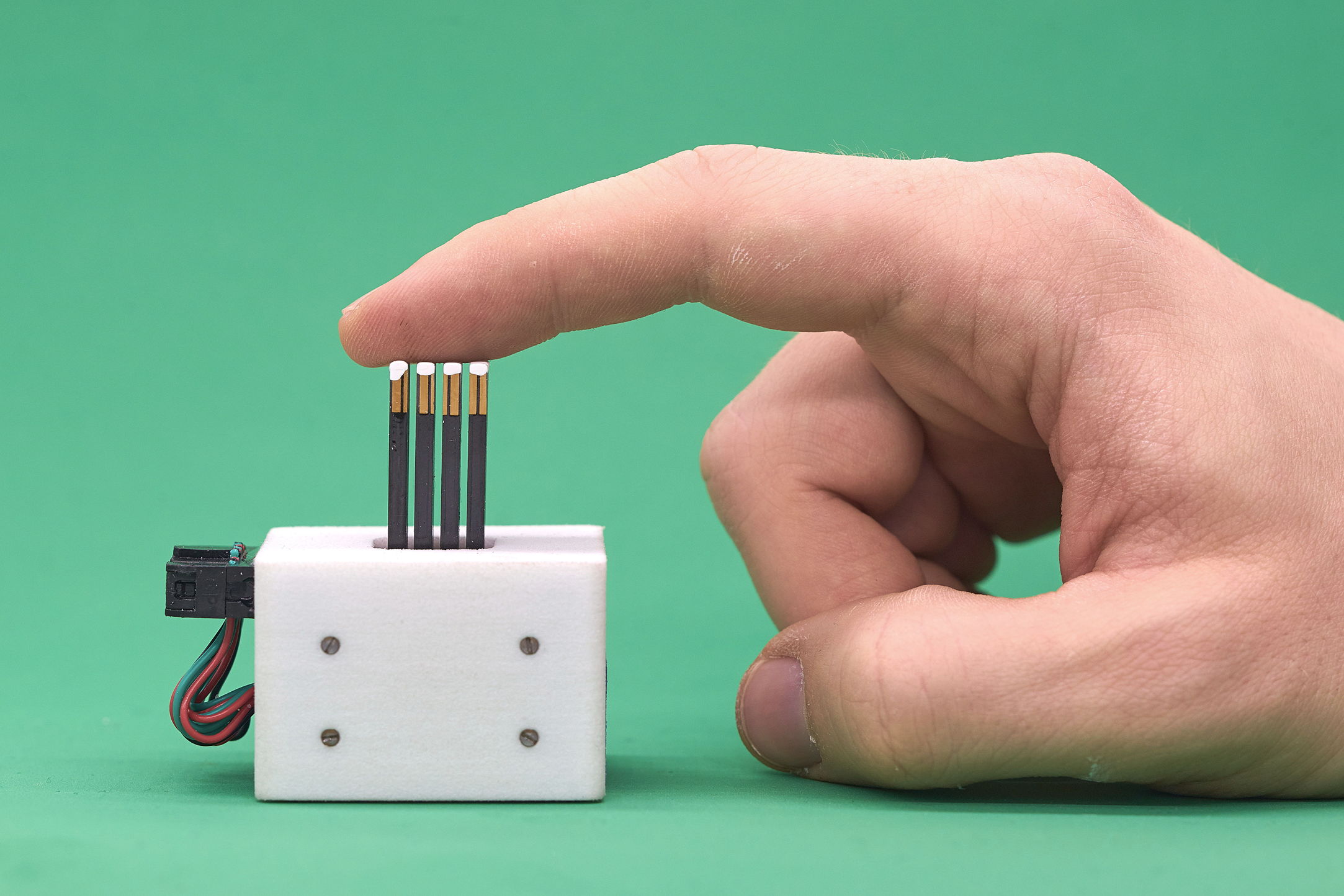}
\caption{3D model of the tactile display and the assembled prototype made from 3D printed parts.}
\label{fig:displayrendering}
\end{figure*}

Since the bearing of the actuators has a rather large impact on the performance of the display, they are vice-like fixated using two 3D printed jaws made from metal. Both jaws are then pressed together using a total of four screws as can be seen in \emph{figure \ref{fig:displayrendering}}. This way, the actuators are tightly fixated. An additional spacer made from acrylic plastic assures the correct position of each actuator, resulting in a spatial resolution of $\SI{2}{mm}$ in the horizontal plane. The casing around this display -- which also is 3D printed -- is designed to tightly fit the display mount of the tactile mouse. A major benefit of the used clamping mechanism is its ability to mount the actuators \emph{as given by the manufacturer} without the need for mechanical adjustments through e.~g. cutting or grinding.

For testing purposes, we connected four of each actuators in parallel, forming a ``line-display'' with five degrees of freedom. Since only $\num{10}$ wires are needed to drive the display in this configuration, we were able to use a relatively small and flexible cord. This way, operating the tactile mouse feels just as natural as using a regular computer mouse. However, the tactile display prototype can be easily extended to utilize all of the $\num{20}$ actuators.

\subsection{Position and Orientation sensors}
To track the absolute position as well as rotation of the mouse two ADNS 9800 optical laser sensors have been used that communicate via \gls{spi} with a Teensy 3.2 board -- the same that are also used within the signal boards. The incremental data acquired by both sensors is integrated using a custom software running on the Teensy and is then supplied using the USB-HID protocol to a external computer (such as e.~g. the \gls{mcu}). Using a set of user defined parameters\footnote{We also provide a graphical user interface for changing these values as well as for calibration in general.} one can adjust the filter characteristic of the integration process, thus either focusing on smoothness of the calculated velocities or a fast reaction to subtle movements.

In this manner a sufficiently accurate, metrical prediction of position, velocity as well as rotational position and velocity can be obtained with an update rate of \SI{500}{\hertz} from the tactile mouse. It should be mentioned, that the accuracy of the metrical prediction highly depends on the used mouse pad as well as on a proper calibration of the system. 
In order to allow for a basic user interaction a RGB \gls{led} as well as a classical mouse button have been added. A second mouse button hidden at the front of the mouse allows to reset the integration process or to initiate the calibration procedure.

The current status of the tactile mouse can easily be acquired by using the \texttt{libITCHy} software library that is part of the \emph{OpenTactile} system. It allows to access position and orientation information as well as the velocity, angular velocity and the status of the thumb button. Setting calibration parameters and changing the colour of the \gls{led} is supported as well.

\subsection{Performance of the Tactile Display}
We validated the functionality of the tactile display by combining measurements of the unloaded case with an analytic model to estimate the achievable tip displacements in case the finger is gently pressed on the display. The mechanical simulation of the \gls{pzt} actuators has been performed using the transfer function method as described by [\cite{pestel1963matrix}]. In order to asses the amplitudes in the unloaded case an optical measurement procedure with an approximate accuracy of $\pm\SI{3}{\mu m}$ was carried out for each of the \num{20} actuators. The amplitudes of the actuators have been measured using a sinusoidal signal of $\SI{20}{V}$ amplitude at a series of discreet frequencies ranging from \SI{10}{Hz} to \SI{400}{Hz}.

\begin{figure}
\centering
\includegraphics[width=0.95\columnwidth]{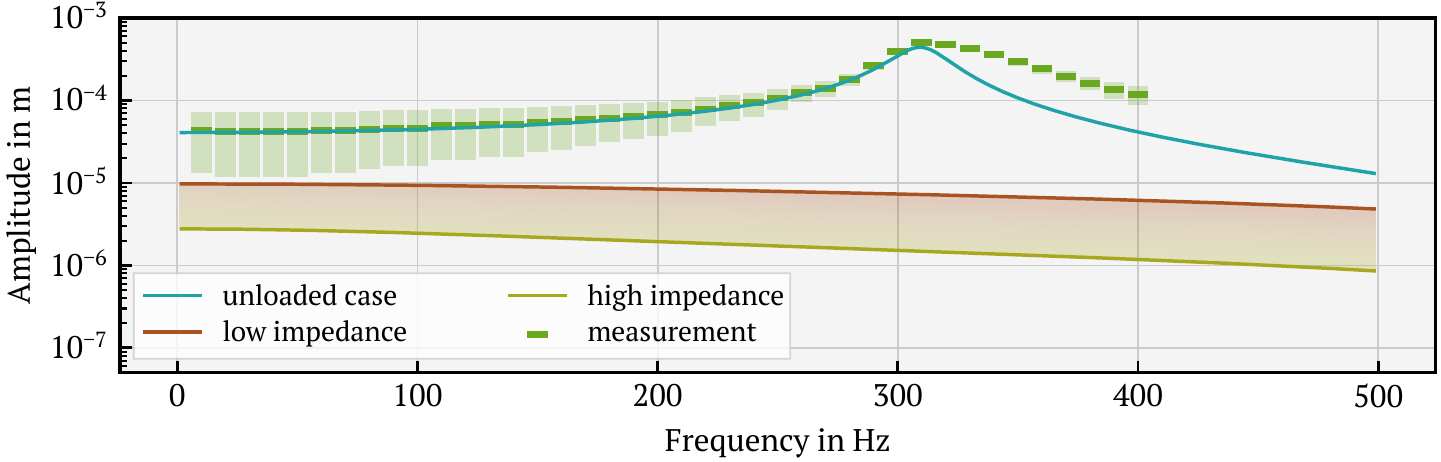}
\caption{Frequency response of an exemplary actuator for a voltage of \SI{20}{V} (green) and the response of the analytic model averaged across all \num{20} actuators (blue). The filled area denotes the range of amplitudes that can be expected in the loaded case of application.}
\label{fig:actuatormodel}
\end{figure}

Given the mechanical and electrical parameters of the \gls{pzt} actuators, an analytic model was fitted numerically to match the measurements -- similar to the procedure described in [\cite{Hofmann2014}]. We assume that these parameters do not vary significantly between individual actuators. Therefore, deviations in mechanical behaviour can be attributed to the mechanical bearing as described in [\cite{Hofmann2017}]. We model this bearing using a simple \emph{Kelvin-Voigt material} that can be described using a total of four free parameters describing the translational impedance ($k_t$, $d_t$) as well as rotational impedance ($k_r$, $d_r$). 
For each of the $i = 1 \ldots 20$ actuators we acquire these parameters by solving a nonlinear minimization problem in a least-squares sense:
\begin{linenomath}\begin{align*}
\argmin_{\substack{k^i_t, k^i_r > \SI{0}{N\per m}\\ d^i_t, d^i_r > \SI{0}{Ns\per m}}} \,\sum_{f = \num{10}\ldots\SI{400}{Hz}} \left\Vert    
   \,\, \hat{u}^{\Gamma_i}_z(2\pi \cdot f) - u^i_f\,
\right\Vert^2 \,.
\end{align*}\end{linenomath}
Here, $u^i_f$ denotes the maximum deflection of the actuator measured at discrete frequencies ($f$), whereas $\hat{u}^{\Gamma_i}_z(\omega)$ is the tip deflection of the analytic \gls{pzt} model for a specific frequency given the bearing parameters ${\Gamma_i} = (k^i_t, k^i_r, d^i_t, d^i_r)$. By calculating the median of these parameters across all measurements, we get a set of ``averaged'' bearing parameters
\begin{linenomath}\begin{align*}
k_t &= \SI{8870}{N\per m},\qquad d_t = \SI{3.62}{Ns\per m},\\k_r &= \SI{0.54}{Nm},\qquad d_r = \SI{1.02e-5}{Nms}\,,
\end{align*}\end{linenomath}
that fit the measurements up to the first resonance peak of the actuator nicely (c.~f. \emph{figure \ref{fig:actuatormodel}}). We suppose the deviations for higher frequencies -- which are present in all measurements around the same frequency range -- to be caused by nonlinearities as well as resonance phenomena of the display housing.

In order to estimate the influence of the fingerpad on the maximum deflections, we used the same procedure as proposed in [\cite{Hofmann2014}]. Here, this influence is modeled as a dynamic load using a \emph{generalized Maxwell model}. The free parameters of this model are fitted against actual measurements as described in [\cite{Hofmann2014}]. Using two such parameter sets -- one denoting the upper impedance limit, the other one the lower limit -- we simulated the influence of this load on the system. \emph{Figure \ref{fig:actuatormodel}} specifies the range of actuator deflections that can be expected when using a driving voltage of $V_{DD} = \SI{20}{V}$. To be able to compare our display with existing approaches, we scaled the results linearly to match the actual voltage we used for testing the system ($V_{DD} = \SI{60}{V}$) as well as the theoretical maximum operating voltage of the used actuators ($V_{DD} = \SI{200}{V}$). These values -- given for the loaded as well as the unloaded case -- are listed in \emph{table \ref{tab:displaycomparison}}. We have to remark, however, that due to nonlinear behaviour of the \gls{pzt} material under higher field strengths (c.~f. [\cite{Wang1999a}]), the actual values may vary.

\begin{figure}
   \centering
    \begin{threeparttable}
    \begin{tabularx}{\linewidth}{p{1.5cm}rrr}
    \toprule
    \multirow{2}{2cm}{Display} & \multicolumn{3}{l}{Amplitude ({\color{Gray} unloaded} / loaded)}                       \\
                                              & $\numrange[range-phrase = -]{0}{50}\si{Hz}$ & $\numrange[range-phrase = -]{50}{300}\si{Hz}$ & $\numrange[range-phrase = -]{300}{1000}\si{Hz}$  \\ \midrule
    \multirow{2}{*}{$\SI{60}{V}$\tnote{\c1}} & \color{Gray} $\SI{123}{\mu m}$     & \color{Gray} $\SI{244}{\mu m}$       & \color{Gray} $\SI{92}{\mu m}$  \\
                                              & $\SI{29}{\mu m}$      & $\SI{26}{\mu m}$        & $\SI{16}{\mu m}$  \\
    \multirow{2}{*}{$\SI{200}{V}$\tnote{\c1}}                         & \color{Gray}$\SI{411}{\mu m}$     & \color{Gray}$\SI{814}{\mu m}$       & \color{Gray}$\SI{305}{\mu m}$ \\
                                              & $\SI{97}{\mu m}$      & $\SI{86}{\mu m}$        & $\SI{53}{\mu m}$  \\ \midrule
    \multirow{2}{*}{Haptex}  & \color{Gray}$\SI{30}{\mu m}$      & \color{Gray}$\SI{33}{\mu m}$        & \color{Gray} -- \tnote{\c2} \\
                                              & $\SI{8}{\mu m}$       & $\SI{6}{\mu m}$         & -- \tnote{\c2} \\
    \multirow{2}{*}{Stress\textsuperscript{2}}  & \color{Gray}$\SI{100}{\mu m}$ & \color{Gray}$\SI{92}{\mu m}$ & \color{Gray}$\SI{54}{\mu m}$ \\
                                              & $\SI{25}{\mu m}$\tnote{\c3} & $\SI{23}{\mu m}$\tnote{\c3}       & $\SI{14}{\mu m}$\tnote{\c3}  \\ 
                         \bottomrule
    \end{tabularx}
    \begin{tablenotes}\footnotesize 
    \item[\c1] Linear projection from measured (\SI{20}{V}) data.\\ Deviations due to nonlinearities have to be expected.
    \item[\c2] No measurements available.
    \item[\c3] Rough guess based on unloaded amplitudes\\(assumed ratio: $1:4$).
    \end{tablenotes}
    \end{threeparttable}
    \caption{Comparison of the proposed tactile Display with two other, \gls{pzt} based approaches. The amplitudes for the ``loaded'' condition are taken from the ``low impedance'' simulation case. The results are compared with published data of [\cite{Allerkamp2007}] and [\cite{Wang2006}].}
    \label{tab:displaycomparison}
\end{figure}

\section{Software Framework and Testing Environment}
Besides the already described software libraries \texttt{libSCRATCHy} and \texttt{libITCHy}, we provide a number of example projects that can be used as a launchpad for creating new applications. In this section we will give a short introduction on using \texttt{libSCRATCHy} by discussing a minimum example program.

We will pay special attention to \emph{ScratchyShow}, a graphical user interface based upon \texttt{libSCRATCHy} that can be used to visualize complex tactile scenarios that can be explored freely using the tactile mouse. The integrated logging functions make this application a powerful platform for designing and conducting user studies.

\emph{This section only contains a short synopsis of the framework and its usage. Information on how to set up the system, API description, etc. can be accessed online.}

\subsection{Basic example using the C++ API}
To give a short introduction in using \texttt{libSCRATCHy}, we will have a look at the minimal example program shown in \emph{listing \ref{lst:code}} that is already sufficient for driving a tactile display in conjunction with the tactile mouse.

\begin{lstfloat*}
\begin{multicols}{2}
\begin{lstlisting}[language=C++]
#include <itchy>
#include <scratchy>

int main(int argc, char* argv[]) {
  GraphicalDisplay display;
  display.show(Icon::Logo, 
               "Demo", "Minimum example");  
  display.detach(); 
    
  FrequencyTable data;
  TactileMouseQuery itchy;
  itchy.initialize();
  SignalManager controller;        
  controller.initializeBoards();
    
  while(true) {
    itchy.update();
    float velocity = itchy.velocity().length();
    
    // Scale frequency with movement speed:
    data.frequency[0] = velocity * 25.0; 
    data.amplitude[0] = 0.9;
        
    for(SignalGenerator& g: 
                controller.generators())
      g.send(data);
    
    display.show(Icon::Logo, "Frequency", 
                 data.frequency[0]);
        
    // Exit loop if button is pressed
    if(display.isPressed(Button::Back)) 
        break;            
  }

  return 0;
}
\end{lstlisting}

\begin{lstlisting}[language=Python,numbers=none,showlines=true]
from SCRATCHPy import *
from ITCHPy import *


display = GraphicalDisplay()
display.show(Icon.Logo, 
             'Demo', 'Minimum example')
display.detach()

data = FrequencyTable()
itchy = TactileMouseQuery()
itchy.initialize()
controller = SignalManager()
controller.initializeBoards()

while True:
    itchy.update()
    velocity = itchy.velocity().length();

    # Scale frequency with movement speed:
    data.frequency[0].value = velocity * 25.0;
    data.amplitude[0].value = 0.9;


    for g in controller.generators():
        g.send(data);

    display.show(Icon.Logo, 'Frequency',
                 data.frequency[0].value);

    # Exit loop if button is pressed
    if display.isPressed(Button.Back):
        break
        
        
                
        
\end{lstlisting}
\end{multicols}
\vspace{-10mm}
\caption{Minimal examples for using \texttt{libSCRATCHy} as well as \texttt{libITCHy}. The left side shows the C++ sources whereas the right side represents the equivalent Python code.}\label{lst:code}
\end{lstfloat*}

Within lines $5$ to $8$, the OLED display is initialized and a simple message is displayed, consisting of one of the predefined icons as well as two lines of text. Since the \gls{i2c} communication with the displays is a blocking process -- and therefore would slow down the execution of the main loop in line $23$ --, the display can be ``detached'' to run in a separate thread.

The subsequent lines $10$ to $13$ are needed for a basic initialization of the system: At first an empty frequency table is declared that later on can be modified and dispatched to the signal boards. Then the tactile mouse and a \texttt{SignalManager} are instanced. By calling \texttt{initializeBoards()} all connected signal boards are automatically enumerated and initialized using default configuration parameters.

After the initialization has been carried out, the signal generation loop is almost self explanatory:
At first, the current status of the tactile mouse is obtained via USB by calling its \texttt{update()} method. Then the velocity magnitude is calculated and afterwards used to modify the first component of the frequency table. The very same frequency table is then send to all signal boards (\emph{lines $23-24$}) and are immediately applied to the connected actuators.

The application is terminated once the user presses the corresponding button on the display breakout (\emph{lines $30-31$}).

\subsection{Python bindings}
While the classical C++ API provides the most flexibility at high execution speed, prototyping more complex models can be cumbersome. Therefore, to allow for rapid prototyping -- and also to relive novice programmers -- we added a Python wrapper to \texttt{libSCRATCHy} that gives access to the highlevel API. The corresponding example code is very similar and can also be seen in \emph{listing \ref{lst:code}}.

One of the main advantages of using the Python API is the similarity to \emph{Matlab} when using additional libraries as \emph{Numpy} and \emph{Scipy}.

\subsection{Designing user studies with ScratchyShow}
To simplify the design of user studies that are carried out using the system, we provide a graphical user interface that is based upon \texttt{libSCRATCHy} and \texttt{libITCHy}. The key idea of this software is, that the user only has to provide the implementation of a model to drive the display, i.~e. a method that adjusts the frequency tables based on the current position and velocities of the actuators. 

Using a special ``scene description'' file format, multiple scenarios can be defined that can be used e.~g. to conduct multi-parted user studies. Within such a scenario, multiple areas can be described that represent e.~g. a specific model or a virtual surface. These areas can be placed freely or in a user definable random fashion across the screen. Each of these surfaces can be given an individual graphical representation, depicting e.~g. the surface structure it is meant to reproduce.

In order to ease the realization of randomized studies, a logging system has been implemented that keeps track of user actions (e.~g. using the thumb button or the average length of stay for each surface and averaged movement velocities) as well as the random choices the system has made. In conjunction with the possibility to replace the surface images with neutral depictions, double-blind studies can be realized easily.

\emph{ScratchyShow is hosted in a separate repository on GitHub. Following the URL}

\url{https://github.com/OpenTactile/ScratchyShow} 

\emph{additional information can be found. This includes a more complete description of the functionality, a specification of the ``scene description'' file format and a short tutorial on how to create own models for driving the tactile display.}

\section{Discussion and Conclusion}
\begin{figure}
    \centering
    \includegraphics[width=\linewidth]{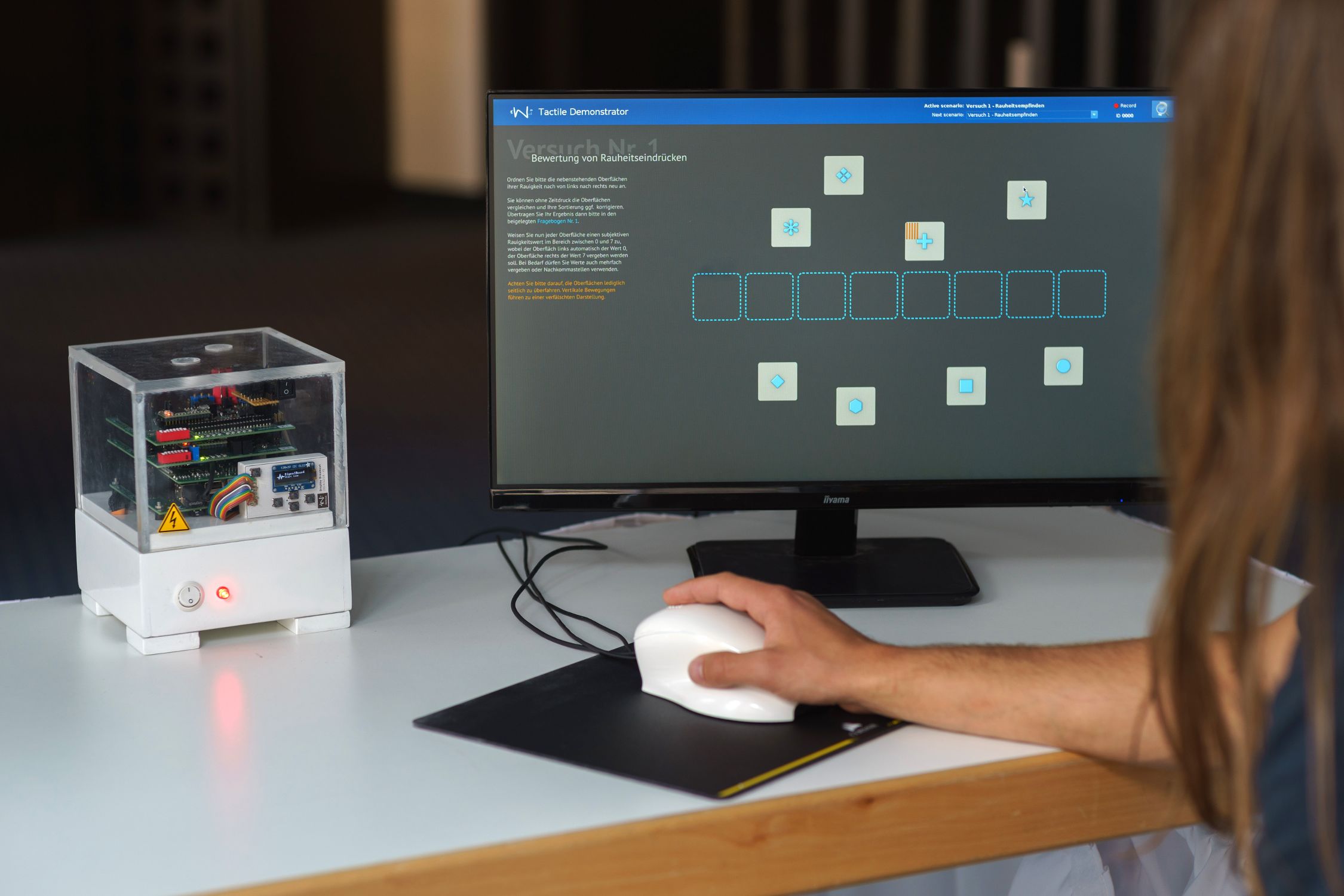}
    \caption{Photograph of the final demonstrator consisting of the tactile mouse and the SCRATCHy hardware system. The ScratchyShow application running on the \gls{mcu} is displayed on the screen.}
    \label{fig:system}
\end{figure}

In this paper, an open source hardware-system has been presented that also includes a software architecture aiming for easy and fast prototyping of tactile applications. The so called SCRATCHy system, as well as ITCHy -- the tactile mouse --, has been verified technically regarding signal quality, latencies and mechanical properties. Beyond the scope of this paper, we successfully used the system throughout various tests and internal studies, using different kinds of tactile displays (see \emph{figure  \ref{fig:system}} for a typical study setup). We therefore are confident that other research groups can benefit from the \emph{OpenTactile} framework as well.

However, the system is currently under active development and there is still much room for improvement. For example, one could tackle possible issues with signal quality of the bus system.
Especially when working with a large amount of signal generators the signal quality as well as electromagnetic compatibility might become an issue. These problems may be engaged from two sides: First and foremost the electromagnetic compatibility should be improved for each of the designed \glspl{pcb}. Additionally a software-based error correction could be implemented.

It should further be noted that within the current revision of the signal generator software no phase correction has been implemented. However, this can be -- if needed -- done with little effort. By including phase information, one could also compensate for low-pass filter induced phase shifts using e.~g. a set of lookup tables integrated into the signal boards. An amplitude compensation, to linearize the filter response, could also be done either on the signal generator or the \gls{mcu}. 

We are currently exploring the option of implementing more sophisticated driving and control schemes. An interesting approach to compensate nonlinear behaviour when driving \gls{pzt} actuators is a charge control as suggested by [\cite{Bazghaleh2010}]. It could be implemented by adding a current measurement unit to each amplifier channel.
In future publications, we will also present more advanced models for driving tactile displays, that can be used in conjunction with this system.

\section*{Funding}
The work was supported by the German Research Foundation (DFG)  (grant number 224743555).

\bibliographystyle{apalike}
\bibliography{library}

\end{document}